\documentclass{WileyMSP-template}
\usepackage{bm}
\usepackage{mathtools}
\usepackage{hyperref}
\usepackage{amssymb}
\newcommand{\MoS}[1]{MoS\textsubscript{#1}}
\newcommand{\MoSe}[1]{MoSe\textsubscript{#1}}
\newcommand{\WS}[1]{WS\textsubscript{#1}}
\newcommand{\WSe}[1]{WSe\textsubscript{#1}}
\newcommand{\PtSe}[1]{PtSe\textsubscript{#1}}

\begin{document}
\title{Atomic Layer-controlled Nonlinear Terahertz Valleytronics in \\Dirac Semi-metal and Semiconductor \PtSe2}
\maketitle
\justifying
\\
\author{Minoosh Hemmat,}
\author{Sabrine Ayari,}
\author{Martin Mi\v{c}ica,}
\author{Hadrien Vergnet,}
\author{Guo Shasha,}
\author{Mehdi Arfaoui,}
\author{Xuechao Yu,}
\author{Daniel Vala,}
\author{Adrien Wright,}
\author{Kamil Postava,}
\author{Juliette Mangeney,}
\author{Francesca Carosella,}
\author{Sihem Jaziri,}
\author{Qi Jie Wang,}
\author{Liu Zheng,}
\author{Jérôme Tignon,}
\author{Robson Ferreira,}
\author{Emmanuel Baudin,}
\author{Sukhdeep Dhillon*}
\dedication{}

\begin{affiliations}
\noindent M. Hemmat, S. Ayari, M. Mi\v{c}ica, H. Vergnet, J. Mangeney, F. Carosella, J. Tignon, R. Ferreira, E. Baudin, S. Dhillon\\
Laboratoire de Physique de l'Ecole normale sup\'erieure, ENS, Universit\'e
PSL, CNRS, Sorbonne Universit\'e, Universit\'e de Paris, 24 rue Lhomond, 75005 Paris, France\\
Email Address:
sukhdeep.dhillon@phys.ens.fr\\
M. Arfaoui, S. Jaziri\\
Laboratoire de Physique de la Matière Condensée, Département de Physique, Faculté des Sciences de Tunis, Université Tunis El Manar, Campus Universitaire 1060 Tunis, Tunisia\\
G. Shash, L. Zheng\\
School of Materials Science and Engineering, Nanyang Technological University, Singapore, Singapore\\
X. Yu, Q. Wang\\
School of Electrical and Electronic Engineering \& School of Physical and Mathematical Sciences, The Photonics Institute, Nanyang Technological University, 50 Nanyang Avenue, Singapore, 639798, Singapore\\
D. Vala, K. Postava\\
IT4Innovations, National Supercomputing Center, VSB -- Technical University of Ostrava, 17. listopadu 2172/15, 708~00 Ostrava-Poruba, Czech Republic
Faculty of Materials Science and Technology, VSB -- Technical University of Ostrava, 17. listopadu 2172/15, 708~00 Ostrava-Poruba, Czech Republic\\


\end{affiliations}



\begin{abstract}
\noindent Platinum diselenide (\PtSe2) is a promising two-dimensional (2D) material for the terahertz (THz) range as, unlike other transition metal dichalcogenides (TMDs), its bandgap can be uniquely tuned from a semiconductor in the near-infrared to a semimetal with the number of atomic layers. This gives the material unique THz photonic properties that can be layer-engineered. Here, we demonstrate that a controlled THz nonlinearity - tuned from monolayer to bulk \PtSe2 - can be realised in wafer size polycrystalline \PtSe2 through the generation of ultrafast photocurrents and the engineering of the bandstructure valleys. This is combined with the \PtSe2 layer interaction with the substrate for a broken material centro-symmetry permitting a second order nonlinearity. Further, we show layer dependent circular dichroism, where the sign of the ultrafast currents and hence the phase of the emitted THz pulse can be controlled through the excitation of different bandstructure valleys. In particular, we show that a semimetal has a strong dichroism that is absent in the monolayer and few layer semiconducting limit. The microscopic origins of this TMD bandstructure engineering is highlighted through detailed DFT simulations, and shows the circular dichroism can be controlled when \PtSe2 becomes a semimetal and when the K-valleys can be excited. As well as showing that \PtSe2 is a promising material for THz generation through layer controlled optical nonlinearities, this work opens up new class of circular dichroism materials beyond the monolayer limit that has been the case of traditional TMDs, and impacting a range of domains from THz valleytronics, THz spintronics to harmonic generation.
\end{abstract}




\section{Introduction} \label{sec:Intro}
A wide breadth of 2D transition metal dichalcogenides (TMD) semiconductors have been investigated in recent years, each with unique bandstructures with properties that are distinct from the bulk~\cite{Manzeli2017,Tan20176225}. Some of the most commonly studied materials include \MoS2, \WS2, \MoSe2, and \WSe2, and key features include tuneable bandgap energies in the optical region in the monolayer (ML) or few monolayer regime, as well as giant exciton binding energies. These materials have shown great potential to impact a range of fields in photonics, from quantum optics~\cite{Tan2020,Xia2014899,GonzalezMarin2019,LopezSanchez2013497}, the emergence of valleytronics \cite{Liu20192695} to layered controlled optical nonlinearities~\cite{Wen2019317}. For example, second harmonic generation has been shown to depend on the number of TMD layers, where an even number of stacked layers possess inversion symmetry whilst an odd number do not \cite{li_probing_2013,kumar_second_2013}. Recently these advances have motivated the application of TMDs to the terahertz (THz) spectral range, historically technologically underdeveloped and where 2D TMDs hold great nascent promise. For example, ML \MoS2 has been been demonstrated as a non-volatile switch for THz communications~\cite{kim_monolayer_2022,kim_zero-static_2018} or as efficient ultrafast THz modulators~\cite{Tan_2DOpto}. THz emission time domain spectroscopy (TDS) has proven to be ideal to probe the relevant ultrafast currents and nonlinearities as a sensitive and non-contact technique ~\cite{Schleicher2009}. Indeed, recent work has shown that optical interband excitations can generate ultrafast THz photocurrents and subsequent nonlinear THz generation in the semiconductors \MoSe2 \cite{Fan202048161,Yagodkin2021} and \WSe2 \cite{Si2018416}. This has permitted novel insights into the various transport phenomena such as photogalvanic effects to shift-currents in these 2D semiconductor materials. THz emission TDS has also shown valuable insights into ultrafast transport in 'zero' bandgap Dirac and Weyl materials. For example, the semi-metal TaAs \cite{gao_chiral_2020}, the Topological Insulator Bi$_2$Se$_3$ \cite{braun_ultrafast_2016} and graphene \cite{maysonnave_terahertz_2014} have shown ultrafast currents that can be controlled through the polarisation of the light excitation. Furthermore, semimetals have shown an important application potential as sensitive THz detectors~\cite{zhang_colossal_2022,qiu_photodetectors_2021,hu_terahertz_2023}.

However, the transition from semiconductor to semi-metal and its effect on the THz photocurrents and nonlinearities generated in TMDs has not been investigated in a single material system, as typical TMDs rapidly transit from a direct to indirect gap with thicknesses greater than one ML. On the other hand, the recently discovered TMD \PtSe2 \cite{wang_monolayer_2015} has garnered considerable interest. Together with relatively large electronic mobilities \cite{zhao_high-electron-mobility_2017}, the layer controlled bandgap can be tuned slowly from a semiconductor to a Dirac semi-metal \cite{ansari_quantum_2019}, covering a vast energy scale, from the near-infrared for MLs to the THz region for multilayers. This has permitted, for example, the recent demonstration of atmospheric stable mid-infrared detectors based on 2D materials~\cite{sefidmooye_azar_long-wave_2021}. Prior investigations of the nonlinear properties of \PtSe2 have been restricted to \PtSe2 as a saturable absorber with no demonstrations on the layer dependent nonlinearities~\cite{YANG2020103155}. Very recent studies have investigated THz photocurrents in \PtSe2 but were limited to a single thickness of 2.4~nm~\cite{zhang2022generation} or 50~nm~\cite{cheng2023giant}, limiting the understanding of the effect of the bandstructure. Further, these studies have assumed an entirely centrosymmetric nature and neglected the effect of the substrate. The latter is primordial to consider in 2D materials, such that the origin of the generated photocurrents can be understood in the context of the layer dependent bandstructure. 

The present contribution shows the first observation of layer dependent, polarization and valley selective excitation of ultrafast photocurents in air stable, wafer scale, ML to multilayer \PtSe2 at room temperature. We show amplitude and phase resolved THz emission from this 2D material under optical femtosecond excitation, permitting access to the processes of ultrafast photocurrent generation. We demonstrate that this is a result of a second-order nonlinear response with the THz pulse generated from the ultrafast photocurrents induced by the nonlinear conductivity of \PtSe2. In particular, we show second order contributions from both linear and circular ‘photon drag’ (PDE) and ‘photogalvanic’ (PGE) effects and the critical role of the environment, with the substrate inducing a structural asymmetry in the inherent centrosymmetry of \PtSe2. This is strongly illustrated by THz photocurrents that show an opposite phase change with left and right circular polarization (LCP and RCP) optical excitation for semi-metal multi-layer \PtSe2 but remain the same for the few layer semiconducting case. These results are corroborated theoretically through extensive DFT simulations of the layer dependent bandstructure showing an important circular dichroism (CD) for semimetal \PtSe2 owing to the excitation of opposite valleys and the strong interaction with the substrate. This work shows that CD is not limited to the ML limit as in other TMDs with natural non-centrosymmetry, and can be finely controlled in the novel layer dependent bandstructure of \PtSe2. 
\begin{figure}[ht]
     \includegraphics[width=\linewidth]{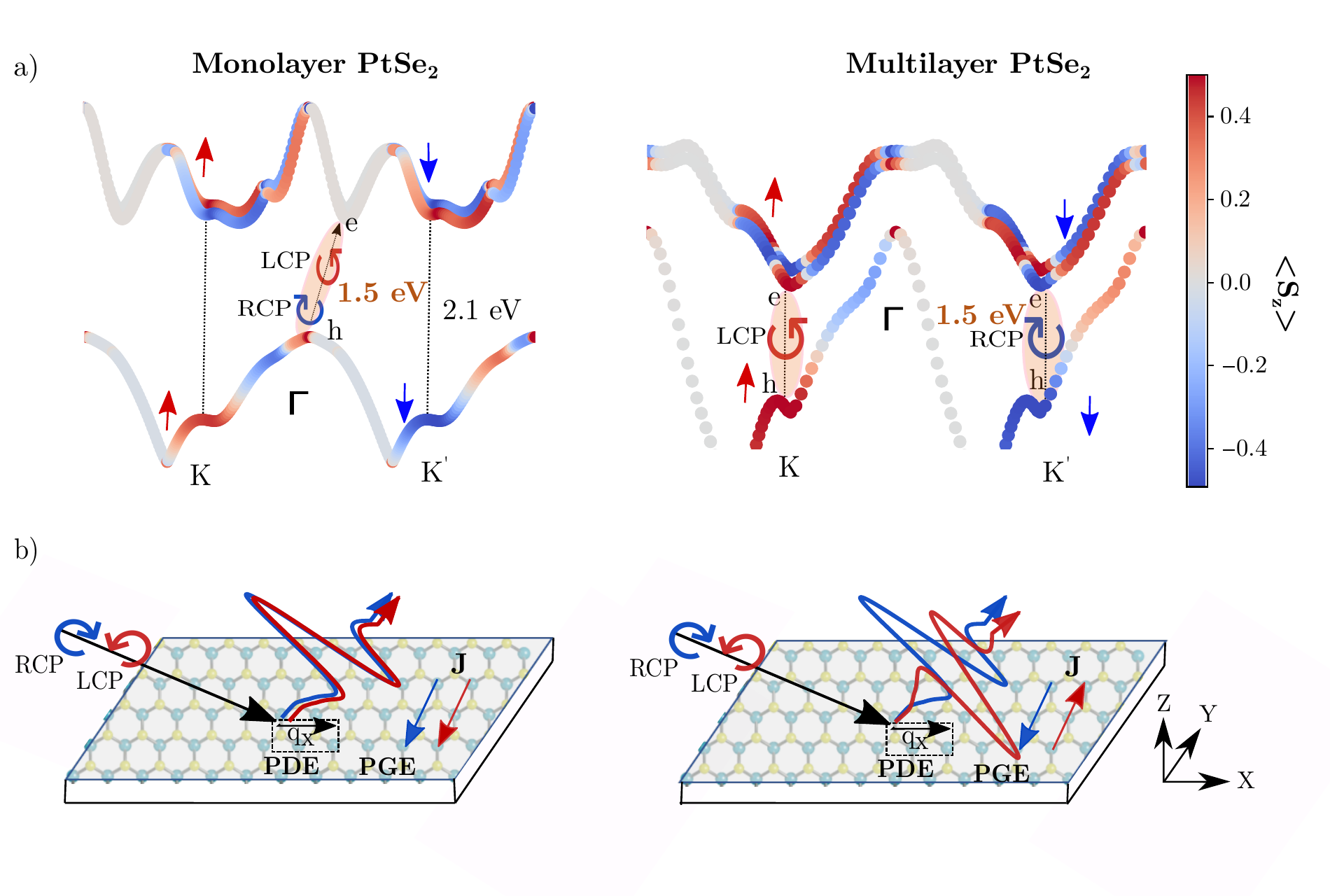}
     \caption{\textbf{Schematic of helicity dependent generation of THz photocurrents in substrate coupled \PtSe2.} a) Simulated spin-resolved band structure (red for spin up and blue for spin down) of ML (left) and multilayer (right) \PtSe2 on asubstrate under circular right (RCP) and left (LCP) femtosecond excitation pulse at 1.5 eV. For the ML indirect interband excitations around $\Gamma$ point (orange) are possible, whilst for multilayer \PtSe2, direct interband transitions in the vicinity of $K (K^{'})$ under circular left (right) polarized light can be excited. b) Schematic illustration of contributions of circular photon drag (PDE) and photogalvanic effects (PGE) and the generation of THz pulses in ML (left, helicity-independent, currents generated with same sign) and multilayer (right, helicity-dependent, currents generated with opposite sign) \PtSe2 with substrate.}
     \label{fig:valley}
\end{figure}
\section{Layer controlled excitation of PtSe$_2$ valleys}\label{sec:Results}
Figure~\ref{fig:valley}a shows the principle of our approach where we exploit the layer dependent energy gaps and the valley properties of \PtSe2. It shows the simulated spin-resolved bandstructure for the two extremes of ML (left) and multilayer (right) \PtSe2 on a substrate. The \PtSe2 is excited with an optical pulse at 1.5~eV, generating ultrafast THz photocurrents that radiate as a short THz pulse. In the ML case, the states at the K points are not accessible for photon energies of 1.5~eV and the interband response is governed by (indirect) transitions around the $\Gamma$ point. Here the bands have mixed spin up - spin down character, thus the optical absorption is not selective to a specific circular polarisation (LCP or RCP) of the incident light. However, this situation changes drastically with multilayer \PtSe2 since the material becomes semi-metallic with increasing of the number of layers (typically greater than a few monolayers) and the interband transitions at the K points become accessible for photon energies of 1.5~eV, dominating the optical response. Through symmetry breaking owing to the substrate, $K$ and $K^{'}$ points have a strong and opposite spin nature that results in a strong CD in \PtSe2 i.e. interband transitions in the vicinity of the $K (K^{'})$ points couple exclusively to LCP (RCP) light. We show that this results in the generation of THz photocurrents and the resulting THz emission with opposite signs when excited with LCP or RCP light. This is not possible in the case of ML semiconducting \PtSe2. This is schematically represented in Figure~\ref{fig:valley}b showing the presence of PDEs and PGEs and the polarisation role on the directionality of the generated current in the $y$ direction ($q_x$ is the photon momentum).

In this work, large area polycrystalline \PtSe2 thin films grown by thermally assisted conversion (TAC) of platinum~\cite{Wang20154013,Wang2019,Gatensby201639,Yim20169550} were studied. These type of films were chosen as the large uniform areas facilate THz investigations (owing to the large wavelengths of THz photons) and where exfoliated samples of any size are notoriously difficult to fabricate. Samples were realised with a range of thicknesses from a monolayer semiconductor to multilayer semi-metal. (0.55, 1.1, 2.2, 3.85, 5.5, 11, 16.5, 27.5 and 38.5 nm). The samples were characterized using Raman microscopy and atomic force microscopy (AFM) to confirm the quality of the samples as well as measuring film thicknesses.. (See supplementary material for further details).
\begin{figure}
    \includegraphics[width=\linewidth]{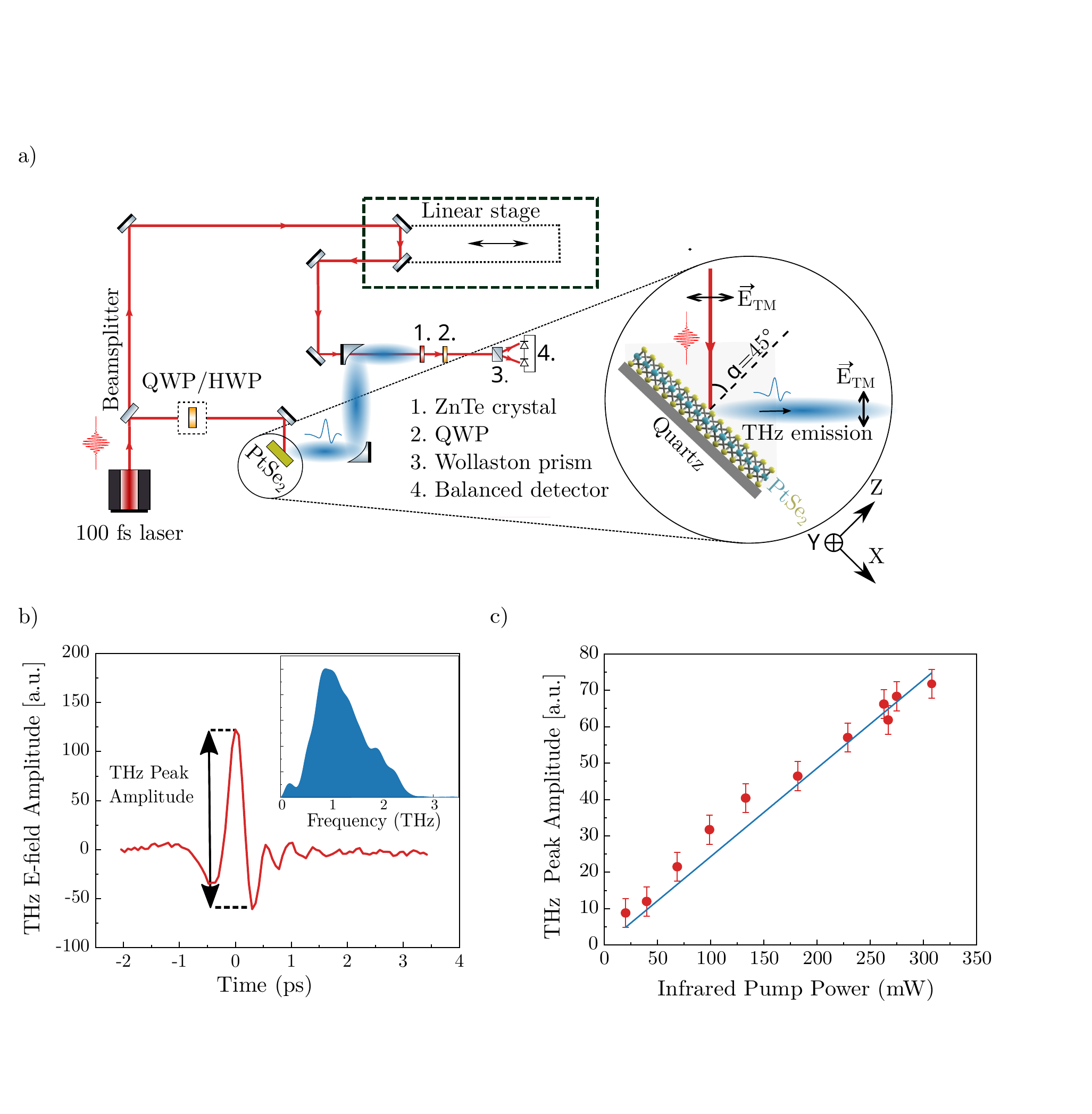}
     \hfil
     \caption{\textbf{Ultrafast THz nonlinear emission from semimetal \PtSe2.} a) Schematic of the experimental setup for THz emission using THz time domain spectroscopy. The magnified schematic illustrates the geometry for THz radiation generation from \PtSe2 sample in reflection configuration. A 100~fs laser Ti:Sapphire oscillator excites the \PtSe2 and induces an ultrafast photocurrent and hence a free space THz pulse, which is consequently detected by EO sampling. b) TM polarised Electric field of single cycle THz pulse as a function of time generated through femtosecond linear (TM) polarisation excitation of multilayer (38.5~nm) semimetal \PtSe2. Fourier transformed spectra of THz electric field is shown in the inset of (b). c) Peak-to-peak THz electric field amplitude dependence on incident average pump power for the 38.5 nm semimetal \PtSe2. The red circles are the experimental data and the blue line is a linear fit.}
     \label{fig:sample}
     
\end{figure}

\subsection{THz nonlinear emission from \PtSe2} 
To investigate the ultrafast photogenerated currents, THz emission TDS with a 100 femtosecond Ti:Sapphire oscillator was employed. A schematic of the experimental setup is shown in Figure~\ref{fig:sample}a. The laser excites the \PtSe2 at an angle of 45\textdegree, generating an ultrafast current and resulting in the emission in free space of a THz pulse. The generated current induces a radiated electric field of the form
\begin{equation}\label{eq:Ethz}
\mathbf{E}(\mathbf{n},r)=\frac{ik}{c}(\mathbf{n}\wedge\mathbf{J_{\omega}})\wedge\mathbf{n}\frac{e^{ikr}}{r}
\end{equation}
where $\mathbf{n}$ is the propagation direction, $r$ is the distance to the source current, $\mathbf{J_{\omega}}=\int\mathbf{j}(\mathbf{x})d^{3}x$ is the integrated source current, $k$ is the wave-vector of the emitted radiation and $\omega=kc$. The electric field of the generated THz signal is then detected using standard electro-optic (EO) sampling with a 1 mm thick ZnTe crystal. The complete temporal THz field is then constructed using a scanning delay line (see Figure~\ref{fig:sample}a). Importantly, this amplitude and phase resolved technique is sensitive to the direction of the generated photocurrents in a non-contact geometry. A characteristic time-dependent single cycle THz wave pulse in transverse magnetic (TM) polarization (Electric field $E\textsubscript{xz}$) for multilayer semimetal \PtSe2 (38.5 nm thick) is reported in Figure~\ref{fig:sample}b. The excitation average power is 250~mW @ 76~MHz corresponding to a pulse energy of $\rightarrow$ 3.3~nJ and was TM polarized. The corresponding THz spectrum - obtained by Fourier transform of the temporal signal - is represented in the inset of Figure~\ref{fig:sample}b. We observe a spectrally broad emission extending from a few hundred GHz to 2.5~THz (limited by the detection crystal). Figure~\ref{fig:sample}c shows the peak-to-peak THz electric field as a function of the infrared pump average power. The observed linear relation between THz field amplitude and IR pump amplitude highlights that THz emission is a result of a second order nonlinear photocurrent generation.

\subsection{Polarization dependency of THz signal}
\noindent To elucidate the actual physical mechanism for the THz pulse in \PtSe2, we investigated the role of the polarization of the optical excitation and the angle of incidence on the polarization of the generated THz pulse, permitting to determine the effects on the generated photocurrents. 
\noindent For any material, the second order nonlinear photocurrent generated from one incident wave $E$ with angular frequency $\omega$ and wavevector $\boldsymbol{q}$ writes~\cite{Ganichev2010}:
\begin{equation}
j_{\lambda}^{(2)}(0,0)=\sigma_{\lambda\nu\eta}^{(2)}(\omega,\boldsymbol{q})E_{\nu}(\boldsymbol{q},\omega)E_{\eta}^{*}(\boldsymbol{q},\omega),
\label{curr}
\end{equation}
where $\sigma_{\lambda\nu\eta}^{(2)}$ is the second-order nonlinear conductivity tensor that can be decomposed up to first order in photon momentum:
\begin{equation}
\sigma_{\lambda\nu\eta}^{(2)}(\omega,\boldsymbol{q})=\sigma_{\lambda\nu\eta}^{(2)}(\omega,0)+\Phi_{\lambda\mu\nu\eta}(\omega)q_{\mu}.
\end{equation}
In a two-dimensional system the index $\lambda$ runs only over the two in-plane coordinates x and y while the indices $\nu$ and $\eta$ can include the normal coordinate z. The first term corresponds to PGEs for which the absorption of a photon results in the appearance of a current, and the second term corresponds to PDEs for which current appears in reaction to the partial transfer of the incoming light momentum. For centrosymmetric materials, such as free standing \PtSe2, PGE are forbidden by symmetry. 
Equation~[\ref{curr}] can be rewritten as the following expressions after separating PGE and PDE photocurrent terms: 
\begin{equation}
j_\lambda^{(P GE)}=\chi_{\lambda \nu \eta}\left\{E_\nu, E_\eta^*\right\}+i \gamma_{\lambda \mu}\left(\vec{E} \wedge \vec{E}^*\right)_\mu 
\label{currPG}
\end{equation}

\begin{equation}
j_\lambda^{(\mathrm{PDE})}=T_{\lambda \mu \nu \eta} q_\mu\left\{E_\nu, E_\eta^*\right\}+i R_{\lambda \mu \nu} q_\mu\left(\vec{E} \wedge \vec{E}^*\right)_\nu    
\label{currPD}
\end{equation}
where $\chi_{\lambda \nu \eta}\ $ and $T_{\lambda \mu \nu \eta}$ are the linear tensor terms while $\gamma_{\lambda \mu}\ $ and $R_{\lambda \mu \nu} $ are circular terms. 

\begin{figure}
    \includegraphics[width=\linewidth]{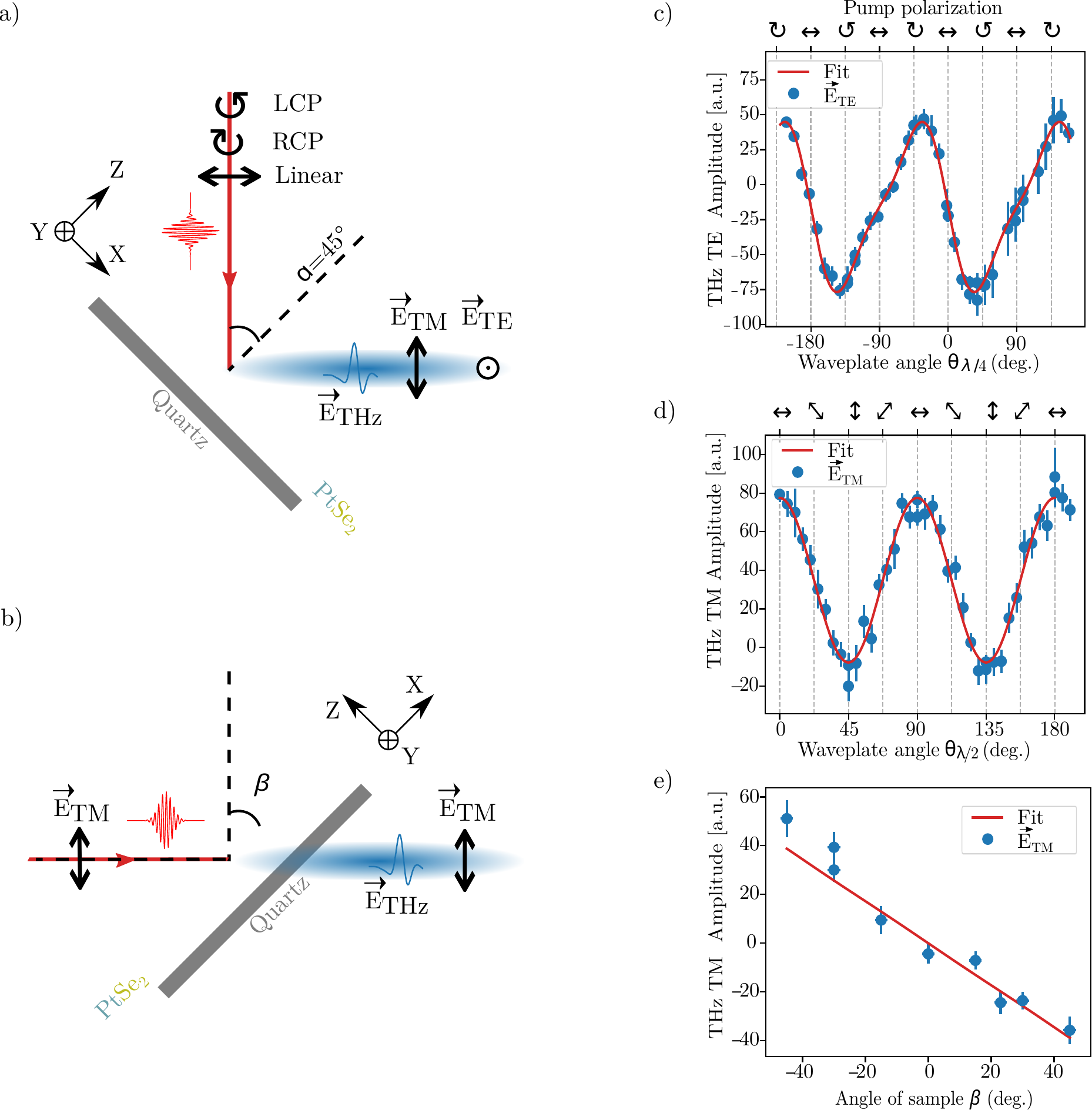}
     \caption{\textbf{Pump polarization and sample angle dependency of the \PtSe2 THz generation - nonlinear tensor determination.} a) Schematic illustration of THz emission TDS in reflection configuration in the presence of a QWP for LCP and RCP or HWP for linear femtosecond excitation. Sample is excited at $\alpha =45^\circ$ and the THz electric field (TE and TM) dependency on the orientation of the pump polarization is measured. b) Schematic of the THz emission spectroscopy in transmission configuration, where $\beta$ is the sample angle where a linear polarized pump beam excites the \PtSe2 and THz-TM emitted pulse is measured. c) THz-TE amplitude as a function of the QWP angle. d) THz-TM amplitude as function of HWP angle. e) Results from transmission configuration where the angle of sample changes in relation to the pump beam and THz-TM amplitude is measured. Blue circles represent experimental results with their error bars and the red line is the fitting. The sample was a 38.5 nm thick \PtSe2 layer}
     \label{fig:carac}
\end{figure}

We present in the following a set of measurements that ultimately leads to the determination of the above tensor components of the nonlinear conductivity and therefore the principal origins of the second order nonlinear photocurrents. It is based on varying the input pump polarisation and its impact on the polarization of the THz beam, which is directly related to the type of photocurrents that are generated. 
This was performed through the introduction of a half waveplate (HWP) or a quarter waveplate (QWP) in the path of the NIR beam as well as analyzing different polarisations of THz signal from the \PtSe2 layers in reflection geometry as shown in Figure~\ref{fig:carac}a. The orthogonal components - TE and TM modes - of THz electric signal were separately measured by using a polypropylene polarizer after the sample where TE is the vertical component of THz electric field ($E\textsubscript{y}$), parallel to the surface of the sample, and TM is the horizontal component ($E\textsubscript{xz}$). In this configuration, while the incident beam angle was fixed at $\alpha =45^\circ$, the circular polarization of the pump beam changed from LCP to RCP by rotating the QWP. The measured TE polarization of THz signal as function of pump polarization angle $\theta$ is shown in Figure~\ref{fig:carac}c. Interestingly we observe a change in sign of the THz field with LCP and RCP, which is discussed further below. The results found to be well fitted to the following phenomenological equation where the THz electric field is calculated after adapting Equation.~[\ref{eq:Ethz}] to the TDS measurement geometry (see supplementary material for a full development).
\begin{equation}
E_{\mathrm{THz}, T E}=(C / 2) \sin \left(4 \theta_{\lambda / 4}\right)+E \sin \left(2 \theta_{\lambda / 4}\right)-D
\label{TE}
\end{equation}
where the 2$\theta_{\lambda /4}$ and the 4$\theta_{\lambda /4}$ terms correspond to the circular and linear contribution, respectively, to the THz signal. The same measurement was performed with a HWP instead of a QWP to investigate the effect of linear polarisation on TM polarised THz emission (Figure~\ref{fig:carac}d). Experimental data showing THz peak-to-peak signal as a function of polarization angle were fitted using the equation:
\begin{equation}
E_{\mathrm{THz}, T M}=A+B \cos \left(4 \theta_{\lambda / 2}\right)
\end{equation}

In these expressions, $D$ is a constant term and is independent of the nonlinear effects while the coefficients $A,B,C$ and $E$ are related to the contributions of both PDE and PGE. In order to separate these effects, measurements were also performed in transmission geometry, where the excitation polarization was set to TM and the THz TM signal change was measured when the sample was rotated (Figure~\ref{fig:carac}b,~e). In this geometry, the incidence and the detection direction were aligned, and the sample orientation (expressed by the $\beta$ angle in Figure~\ref{fig:carac}b) changed. The THz emission electric field was calculated to be the following equation to which the experimental data were fit: 

\begin{equation}
E_{\mathrm{THz}, T M}=\cos (\beta) \sin (\beta-\alpha)\left[F+G\sin ^2(\beta-\alpha)+H\cos (\beta-\alpha)\right]    
\end{equation}
Through the determination of these coefficients from the above fits, we can extract the tensor components of the nonlinear conductivity of equations~[\ref{currPG}] and ~[\ref{currPD}], adapting an approach used for graphene \cite{Jiang_2011}. Further details of each of the following tensors is given in the supplementary material. Extra tensor terms of \PtSe2 average to zero owing to its polycrystalline nature. Interestingly, the terms considered are adapted for polycrystalline materials meaning that they are robust against rotation. The contribution of the linear PDE: $T_{1}$ = $T_{xxxx}+T_{xxyy}$= ${159 \pm 38}$ (m/Vs) , $T_{2}$ = $T_{xxxx}-T_{xxyy}$= $ {162 \pm 38}$ (m/Vs) , $T_{3}$ = 2$T_{xzxz}$= $ {98 \pm 49}$ (m/Vs) and $T_{4}$ = $T_{xxzz}$= ${67 \pm 16}$ (m/Vs) and the linear PGE: $\chi_{xxz}$\ + $\chi_{xzx}\ $= $ {-104 \pm 41} $ (m/Vs), are found to be present and of similar magnitude. The total circular contribution of the PDE and the PGE are smaller than the individual linear parts with $ R_{2}-R_{1}-\sqrt{2}\gamma$ = $R_{xyz}-R_{yzx}-\sqrt{2}\gamma_{xy}$= ${24 \pm 0.02} $. (The individual components of circular PDE and PGE cannot be separated in this analysis). This shows that both PGE and PDE are present in \PtSe2 for the nonlinear THz generation. Importantly the presence of PGE points to the breaking of the \PtSe2 crystal symmetry. Note that as \PtSe2 is polycrystalline, a huge number of grains contribute to the THz emission. However, each grain is expected to be large enough to permit a band structure description, thus individually supporting the existence of a PDE, which is independent of the intrinsic orientation of its unit cell atoms with respect to the substrate. In addition, the interaction with the substrate brings a vertical ($z$) perturbation to each grain, such that the PGE discussed above holds for each grain taken individually. The microscopic origins and the involved states governing these effects are discussed below.
\subsection{\PtSe2 thickness dependence of THz emission - from monolayer to multilayer} 
\begin{figure}
     \includegraphics[width=\linewidth]{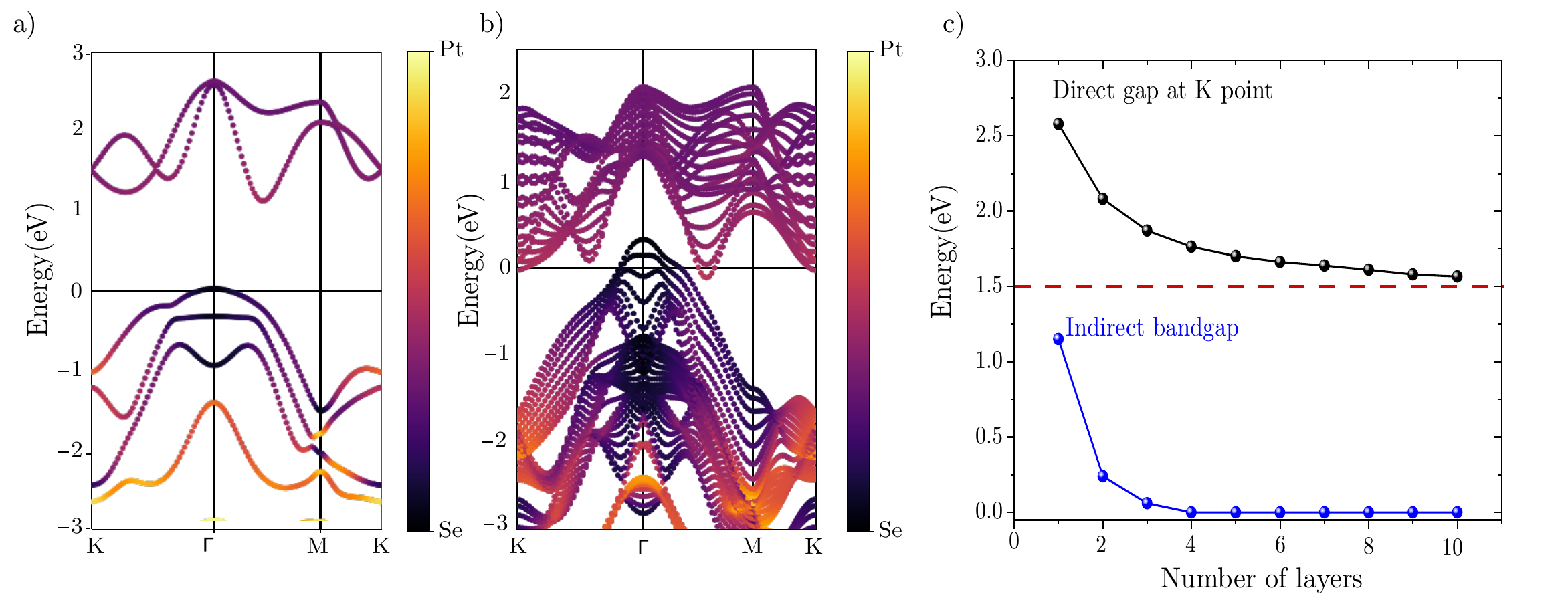}
     \caption{\textbf{Layer dependent Bandstructure of \PtSe2.} \PtSe2 band structure including spin-orbit coupling for a) ML and b) multilayer cases (10 ML).  Energies are referenced to the Fermi level. c) Indirect bandgap at $\Gamma$-M and direct bandgap at $K$ points as a function of number of \PtSe2 layers}
     \label{fig:band}
\end{figure}

 This section presents the strong effect of the semiconductor to semimetal transtion on the THz emission that occurs when increasing the number of MLs. The DFT calculated band structure of the \PtSe2 along the high-symmetry lines of the Brillouin Zone (BZ) for the two extremes of a ML and multilayer structures are shown in Figure~\ref{fig:band}, where spin-orbit coupling is included. This shows that ML PtSe$_{2}$ is a semiconductor with an indirect band gap equal to 1.15 eV (see Experimental Section/Methods), with the valence band maximum (VBM) residing at the $\Gamma$ point and conduction band minimum (CBM) along the $\Gamma-M$ line, which is dominated by $d$ states of Pt and $p$ states of Se (Figure~\ref{fig:band}). As the thickness increases, the bandgap decreases and the electronic structure evolves into a semimetal for $N >4$ (Figure~\ref{fig:band}c and supplementary material). (Reported experimental investigations of this transition with number of layers, however, have shown a variety of results spanning from a few monolayers to tens of monolayers ~\cite{Wang2019,xie2019optical}). This transition originates from the strong electronic interlayer hybridization of the $p_{z}$ orbital of the Se atom (See Figure S4 in the supplementary material). It should be also noted that the energy gap at the K points decreases with the increasing number of \PtSe2 layers, becoming resonant with 1.5 eV for layers $>$ 10 ML, with important consequences for the THz emission properties (see below).

\begin{figure}
     \includegraphics[width=\linewidth]{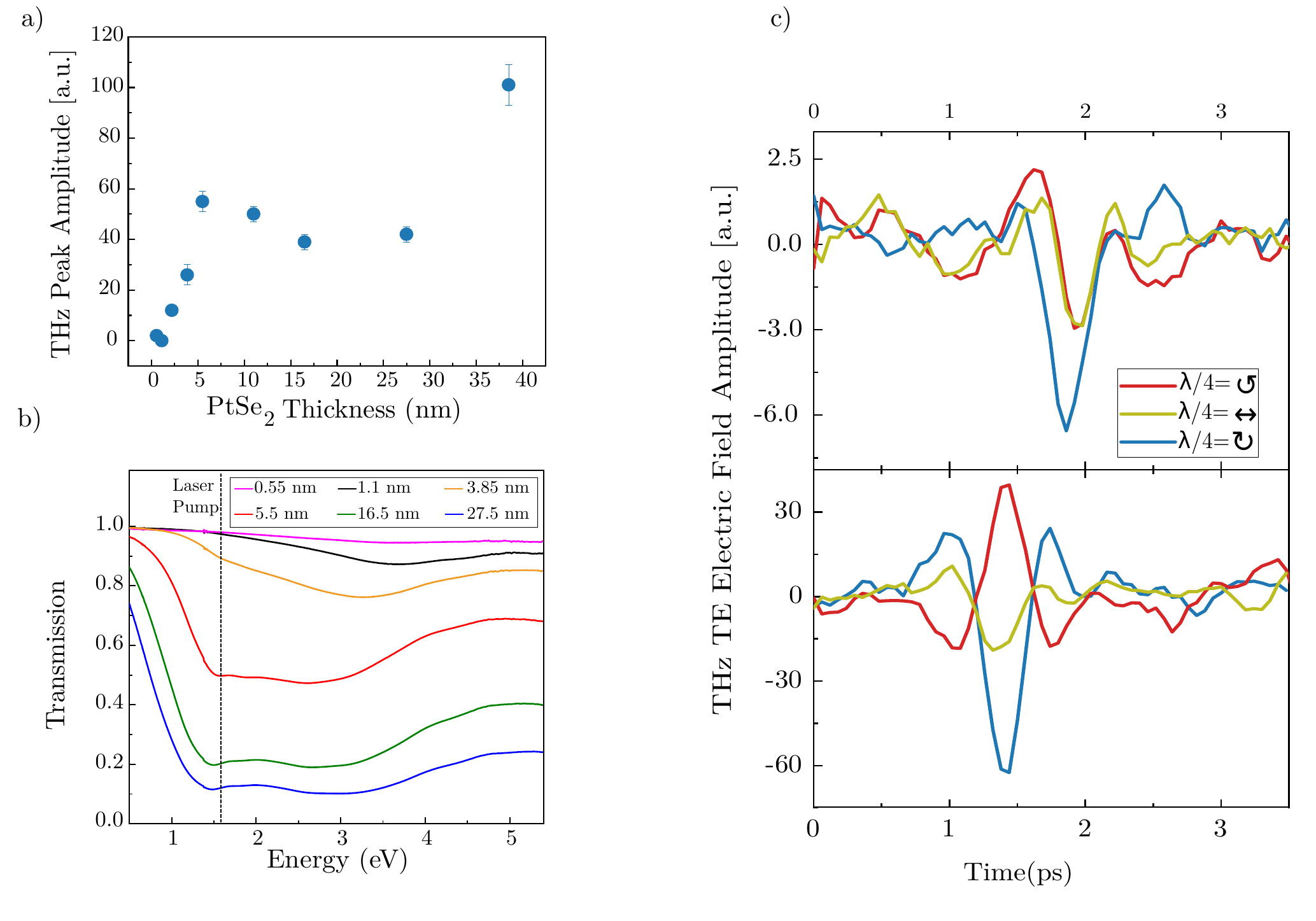}
     \caption{\textbf{Layer controlled THz and optical properties of \PtSe2 - From monolayer to bulk.} a) THz peak amplitude as function of thickness of the \PtSe2 samples.The measurements were done in reflection geometry at 45$^\circ$ with TM polarization of the optical pump and THz TM generated pulse. b) Optical transmission for different \PtSe2 thicknesses extracted over the UV, visible and near infrared spectral regions. c) THz TE electric field of 5.5 nm (top) and 38.5 nm \PtSe2 (down) under linear (TM), LCP and RCP polarization laser excitation.}
     \label{fig:Hellicity}
\end{figure}

Considering these large changes in the bandstructure with number of MLs in \PtSe2, we investigated the THz emission properties in the reflection geometry described above for the range of samples with thicknesses down to one ML. Figure~\ref{fig:Hellicity}a represents the measured peak THz electric field amplitude for different \PtSe2 thicknesses showing a non-trivial response. As the number of layers are increased starting from a ML, the peak field appears to increase linearly with a sharp change around 5.5 nm thick \PtSe2 where the THz field reaches a plateau or starts to drop slowly until 16.5 nm and then starts to increase again. Approximately around the thicknesses of 5.5 nm, multilayer \PtSe2 is expected to have become a semi-metal and the $K$ valleys become accessible to the optical pump. This behaviour suggests that the change in bandstructure has an important impact on the THz generation mechanism. The excitation of the $K$ valleys could also be at the origin of the saturation of the THz field for \PtSe2 films greater than 5.5~nm and discussed further in the microscopic origins section.

To further correlate these observations to changes in bandstruture, the optical transmission of the samples was determined (see Experimental Section/Methods). The results are depicted in Figure~\ref{fig:Hellicity}b where a sharp red shift is observed in the transmission as the thickness is increased beyond 5.5 nm. This sudden modification of the absorption properties reinforces the hypothesis of a transition at around 5.5 nm thick \PtSe2. Similar transmission measurements were performed on \PtSe2 thin films by He et al.~\cite{he2020optical}. Small differences between results can be attributed to different substrate used in this study (quartz instead of sapphire), different sample preparation and quality of the surface. A 5.5 nm thickness corresponds to $\sim$ 10 MLs, and where the K points theoretically begin to be resonant with the 1.5 eV optical excitation. It should be noted that our simulations agree with previous published calculations that predict a faster decrease of the bandgap ($\Gamma$ and $K$ points) with number of layers than in experiments. Therefore the $K$ points are expected to be resonant for 1.5 eV for slightly thicker \PtSe2 layers. This difference is possibly a result in the great sensitivity of the electronic bandstructure to the interlayer distance and coupling, the growth method and/or the strong interaction of these materials with the environment as discussed further below.

\begin{figure}
     \centering
     \includegraphics[width=0.9\linewidth]{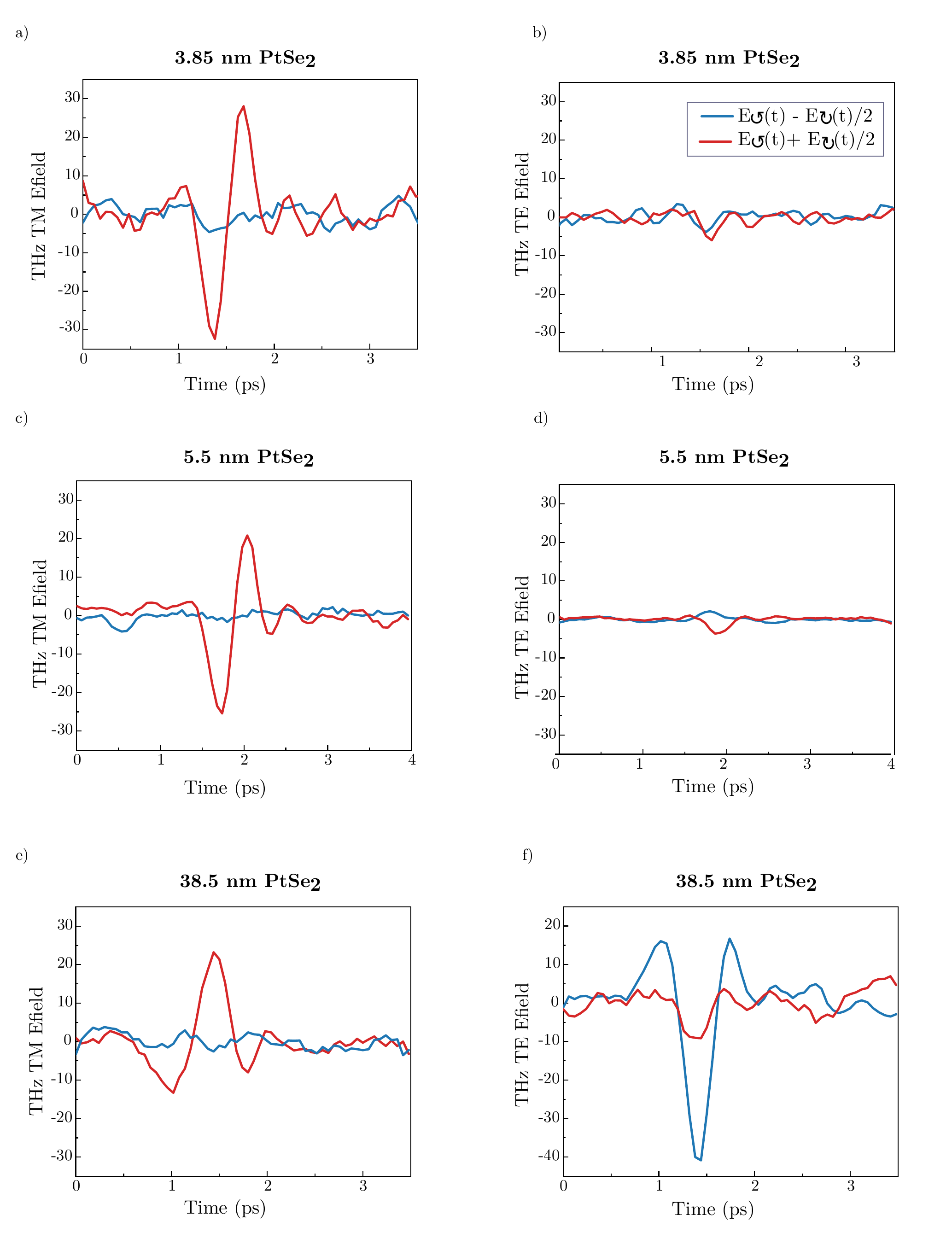}
     \caption{\textbf{Contributions of circular $\bm{(E_{\circlearrowleft}-E_{\circlearrowright})/2}$ and linear $\bm{(E_{\circlearrowleft}+E_{\circlearrowright})/2}$ effects to \PtSe2 THz emission.} Circular and linear effects shown by blue and red lines, respectively, are obtained from measurements in reflection geometry with LCP and RCP excitations. Figures on the left show these effects for THz TM emission and right figures are for the THz TE emission for a,b) 3.85 nm \PtSe2 c,d) 5.5 nm \PtSe2 and e,f) 38.5 nm \PtSe2 at 1.5 eV photon energy.}
     \label{fig:LE-CE}
\end{figure}

This bandstructure flexibility of \PtSe2 with layers and its sensitivity to the environment can also be used to control the properties of the THz pulse and in particular through the asymmetry induced CD when optical interband transitions are excited at the K points. Figure~\ref{fig:Hellicity}c shows the generated TE THz pulse from the few layer 5.5 nm (upper) and multilayer 38.5 nm (lower panel) \PtSe2 layers under LCP (red), RCP (blue) and linear (yellow) polarization. We observe completely different behaviours of the thin and thick samples. For the thick sample the THz signal changes sign with the chirality of the circular polarization, indicating a CD and photocurrents that are generated in opposite directions (see Figure~\ref{fig:carac}c and Equation.~[\ref{TE}]. However, for the thin samples, little dependency on the incoming circular polarization is observed. The 3.8~nm sample showed a similar response to the 5.5 nm thick one (see supplementary material). When the optical transmission of these samples are compared (Figure ~\ref{fig:Hellicity}b), we clearly see that only the thick samples display a permittivity peak at the laser excitation energy (1.5 eV) corresponding to excitation around the $K$ points of the bandstructure. Indeed, the interband transition at the $K$ point is expected at a much higher energy for the few layer samples (at 2.5 eV for the ML case - see Figure~\ref{fig:band}c). To highlight further the observed trends for different \PtSe2 thicknesses under circularly polarized light, the linear effects were extracted as $(E_{\circlearrowleft}+E_{\circlearrowright})/2$ and the circular effects as $(E_{\circlearrowleft}-E_{\circlearrowright})/2$ \cite{ni_giant_2021} from the above plots. These are shown in Figure~\ref{fig:LE-CE} for 3.85~nm, 5.5~nm and 38.5~nm \PtSe2 for both TM (a, c and e) and TE (b, d and f) polarisations. As expected (see supplementary material) only the THz TE signal displays a circular effect. However, it is only observed for the thickest films ($>$ 5.5~nm, despite the linear effects being similar for TM polarisation. This further highlights the strong effect of the \PtSe2 bandstructure on the THz properties. Further, it should be noted that if only PDEs were present, a CD would be expected for thin \PtSe2, showing that PGEs play an important role. 
\subsection{Microscopic origins of circular dichroism in thick \PtSe2}\label{sec:PolCurr}
We have shown above that the THz electric field in \PtSe2 depends strongly on the number of layers, the polarization and angle of the incident beam, and that both PDE and PGE contribute. We now consider the microscopic origins of polarization-dependent photocurrents and the existence of CD related to excitations with opposite photon helcities, LCP and RCP. In particular, we determine the optical matrix elements (OMEs) between the top valence band (VB1) and the bottom  conduction band (CB1) electronic states (see Experimental Section/Methods), which is directly proportional to the interband photocurrent that gives rise to the polarization dependent nonlinear THz emission.  
It should be noted that free standing \PtSe2 possesses centrosymmetric space group $P\bar{3}m1$ for the global structure and polar point groups $C_{3v}$ and $D_{3d}$ for the Se and Pt sites, respectively. This results in a bandstructure that is two-fold degenerate and, for any BZ wavevector, one has two interband transitions for the same photon energy, with same intensity (same absolute value of their OME) but for opposite helicities. This leads to a vanishing CD over the whole BZ as shown in figure S5 in the supplementary material. The presence of an external perturbation, breaking the centrosymmetry, lifts the band degeneracy and triggers a CD effect. Here, the sample deposition breaks the vertical centrosymmetry between the bottom (substrate) and upper (vacuum) regions around the ML. To better understand its effect we consider a toy model whereby a small phenomenological shift is imposed on the position of Se atoms of the lower ML plan to mimic a deposition effect (slightly shifting only their site energies leads to qualitatively same effects). 

\begin{figure}
    \includegraphics[width=\linewidth]{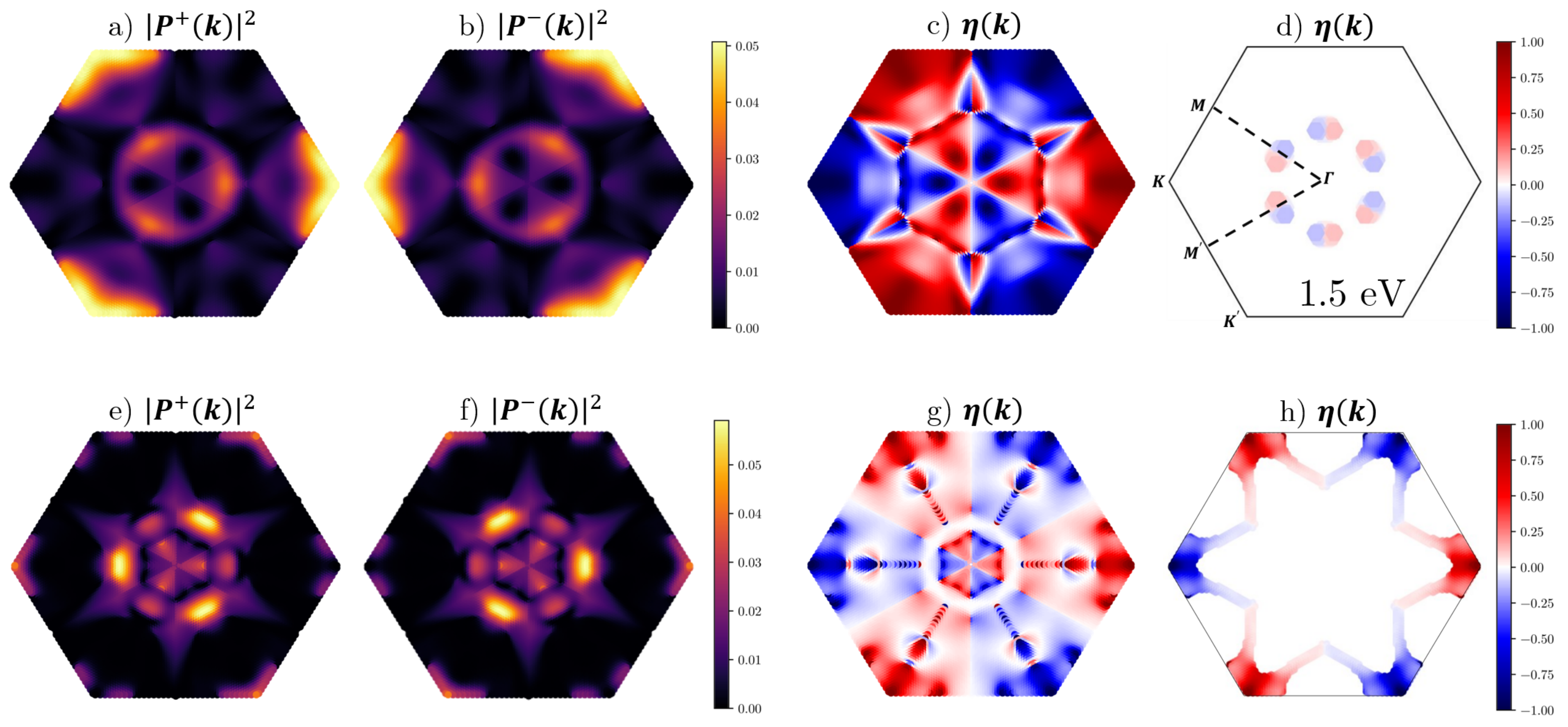}
     \caption{\textbf{Layer and photon energy dependent circular dichroism in \PtSe2.} Momentum dependence of circular polarization of deposited ML (top pannel) and 10 ML (botton pannel) of \PtSe2 between the top of the valence band (VB1)
 and  the bottom of the conduction band (CB1) : (a, b, e, f) the k-resolved direct interband optical matrix elements $\mathcal{P}_{c v}^{\pm}(\mathbf{k})$, (c-g) the degree of optical polarization over all energies $\eta(K,\hbar \omega_{c,v} )$ and (d-h) $\eta(K)$, for a fixed excitation energy ($\hbar \omega_{c,v}=1.5  \pm$~0.06~eV) }
     \label{fig:dicho_ML_bulk}
\end{figure}

Figure~\ref{fig:dicho_ML_bulk} shows the k-resolved direct interband OMEs 
$\mathcal{P}_{c v}^{\pm}(\mathbf{k})=1 / \sqrt{2}\left[P_x^{c v}(\mathbf{k}) \pm i P_y^{cv}(\mathbf{k})\right]$, and the degree of optical polarization $\eta\left(\mathbf{k}, \hbar \omega_{c v}\right)=\frac{\left|\mathcal{P}_{ c v}^{+}(\mathbf{k})\right|^2-\left|\mathcal{P}_{c v}^{-}(\mathbf{k})\right|^2}{\left|\mathcal{P}_{c v}^{+}(\mathbf{k})\right|^2+\left|\mathcal{P}_{ c v}^{-}(\mathbf{k})\right|^2}$ owing to the interband excitation over all energies and energy filtered at 1.5~eV in momentum space for deposited  ML (top - a, b, c, d) and multilayer (bottom panel - e, f, g, h) \PtSe2  for LCP and RCP light ($\sigma_{\pm}$). The energy-filtered CD response around 1.5~eV is obtained by only retaining the contributions of states with $\hbar \omega_{c v}(k) = \hbar \omega \pm$~0.06~eV. The dependence on the transition energy $\hbar \omega_{c v}(\mathbf{k})=\varepsilon_c(\mathbf{k})-\varepsilon_v(\mathbf{k})$ is implicit through k. Figure~\ref{fig:dicho_ML_bulk} shows significant CD in the high-symmetry direction $\Gamma-K$ with $\eta =\pm 1$ and negligible spin polarization near the $\Gamma$ point. This microscopic chiral selection rule arises from the asymmetry of the global crystal space group induced by the substrate and leads to valley-dependent optical selection rules for LCP and RCP excitation in both ML and multilayer \PtSe2 i.e. excitation with LCP or RCP polarized light only leads to an absorption at the $K$ or $K^{'}$ point respectively. However, Figure~\ref{fig:dicho_ML_bulk}(d) shows that ML \PtSe2 results in a complete lack of CD for energies of 1.5~eV around the $K$ points. (The optical transition in the vicinity of the $K$ ($K^{'}$) point requires an excitation of 2.5~eV). As the \PtSe2 thickness is increased beyond the semimetal transition, the CD at the K points increases greatly, as the K points becomes resonant with the optical excitation at 1.5~eV as shown in Figure~\ref{fig:dicho_ML_bulk}(g) for the multilayer structure. Although traditional ML TMDs have already shown to exhibit such strong CD owing their inherent broken inversion symmetry (giving rise to the field of valleytronics), \PtSe2 is distinct in that the CD is not limited to the ML and its inherent centrosymmetric nature. Here the multilayer semimetal \PtSe2 shows a strong valley selective CD through interactions with the substrate, resulting in opposite sign in the generated ultrafast THz photocurrents\cite{xiao_coupled_2012}.

This microscopic approach can also bring insights into the THz field behaviour with \PtSe2 layers (Figure~\ref{fig:Hellicity}a). The deposition perturbation is expected to act on all atoms within a characteristic distance $d_c$ from the substrate (and not only on the low-plan atoms). Thus, increasing the number of layers from 1 ML should lead initially to an increase of the PGE, which would nevertheless saturate for thicknesses bigger than $d_c$. Second, the wavevector region for resonant excitation at 1.5 eV shifts towards the $K$ and $K^{'}$ points with increasing number of layers (Figure~\ref{fig:band}), so that the energy detuning with respect to excitation at these high-symmetry points decreases with increasing sample thickness. Both trends collaborate to an initial increase of the PGE with increasing sample thickness, followed by a plateau in the THz-amplitude signal. This is in qualitative agreement with the measures in Figure~\ref{fig:Hellicity}a. In addition, note that $d_c$ should depend not only on the details of the sample-substrate interactions, but also be affected by the increasing number of free-carries that appear owing to the increasing semi-metallic character of \PtSe2. Indeed, while the direct-to-indirect transition is expected for a few layers, the spectral region where valence and conduction bands overlap slowly increases with increasing number of layers. That means that the density of free carriers in the sample also slowly increases, and as a consequence also the electrostatic screening of the external perturbation. This may explain the slow decrease of the THz amplitude in Figure~\ref{fig:Hellicity}a. The important increase in Figure~\ref{fig:Hellicity}a for very thick samples may be attributed to an increasing PDE contribution. Indeed, the PDE is expected to occur for any sample thickness and not depend on the presence of a symmetry-breaking perturbation. However, as a second-order effect, its strength is substantially attenuated for thin samples because the photon energy is highly non-resonant with respect to the $K$ and $K^{'}$ transition (that dominantes at 1.5 eV in the bulk limit), and becomes more and more resonant as the number of layers increases. In addition, the appearance of an increasing density of free carriers in the sample may also bring a further contribution to the PDE (although non-resonant).

\section{Conclusion}
To conclude, this work has investigated ultrafast photocurrents and the resulting THz nonlinearities in \PtSe2 and highlighted the valley control of these nonlinearities via the number of atomic layers. Unlike other 2D materials, this work shows the important effect of the thickness of this TMD and how the transition from semiconducting to semimetallic behaviour can be used to control the ultrafast photocurrents. Further, we show the potential of valleytronics in thick \PtSe2 layers compared to the ML layer limit for other TMDs. Indeed, by engineering the TMDs thickness such that different $K$ valleys can be excited with a chosen laser excitation and circular polarisation, leads to the control of the photocurrent direction and hence the THz emitted phase. These conclusions are supported by extensive bandstructure simulations that shows the presence of circular dichroism for multilayers ($>$~10~ML) but is absent for few MLs. Vitally, the role of the environment is shown to be crucial to explain our results, which permits to break the centrosymmetric nature of \PtSe2 to induce a giant second order nonlinearity and circular dichroism. As well as understanding and bringing insights into layer controlled nonlinearites and transport phenomena in 2D THz materials, these results are promising for \PtSe2 to be potentially exploited in THz valleytronics, spintronics, harmonic generation and optoelectronic applications.
\section{Experimental Section/Methods}\label{sec:Methods}

\textbf{Materials.} Large areas of \PtSe2 samples were fabricated by thermally assisted conversion (TAC) of Pt metal on a quartz substrate. The thickness of Pt metal varies from 0.1~nm to 7~nm thick. The resulting thickness of \PtSe2 was measured by atomic force microscopy (AFM, see supplementary material) giving a conversion ratio of 5.5 between selenized thin film and the original Pt film. Micro-raman spectroscopy was performed at room temperature with a Renishaw Invia system equipped with a 532~nm laser (see supplementary material). 
\\
\textbf{Experimental setup.} In order to investigate the ultrafast generated photocurrents and THz nonlinearities in \PtSe2 samples, THz emission time-domain spectroscopy (TDS) was performed. A NIR 100-fs pulsed Ti:Sapphire laser source (Coherent Mira 900) with a central wavelength of $\lambda$ = 800 nm (1.5 eV) and repetition rate of 76 MHz was used to characterize coherently the THz emission and photocurrents generated in \PtSe2 samples. The output beam was split in two parts. The intense part (pump) excites the sample to generate a THz pulse while the weak part (probe) passes through a delay line. The average incident power on sample was 250 mW. The generated THz pulse is detected using the probe and a 1 mm-thick ZnTe electro-optic crystal. The electro-optic detection part of the setup consists of a waveplate, Wollaston prism and two balanced photodiodes (Figure~\ref{fig:sample}). In reflection configuration, the pump beam was incident at an angle of 45$^\circ$ relative to the normal of the sample and then focused down to $\sim$ 100 $\mu$m. A half waveplate (HWP) or a quarter waveplate (QWP) was introduced in the pump beam path to change the polarization of the beam and study their effect on THz emission when both incident and detection angle of the pump beam were fixed. The THz beam after the sample, and its transverse electric (TE) or transverse magnetic (TM) polarization was measured separately. In transmission configuration, the polarization of the pump beam was fixed as linear in TM mode. The angle of the pump beam and the detection angle were changed and the THz emission in TM polarization was studied.
Transmission spectra in UV-Vis-NIR region were obtained using an Agilent Cary 7000 double beam spectrophotometer with clean substrate without \PtSe2 layers placed in reference beam path.
\textbf{Theoretical calculations.} The structural relaxation and electronic properties were performed by Quantum Espresso (QE) code with Density functional theory (DFT) based on Generalized Gradient Approximation (GGA) of Perdew-Burke-Ernzerhof (PBE) exchange-correlation functional \cite{giannozzi2009quantum,giannozzi2017advanced,perdew1996k}. Self and non-self consistent DFT calculations are performed to obtain the Kohn-Sham eigenvalues and eigenfunctions.  After obtaining the DFT-KS eigensystem, we use YAMBO code~\cite{sangalli2019many,marini2009yambo} to  calculate the direct  interbrand matrix elements ${P}^{{c} v}(\mathbf{k})=\left\langle\psi_{c \mathbf{k}}|\hat{p}| \psi_{v \mathbf{k}}\right\rangle$. A plane-wave basis set is employed at a cutoff energy 680 eV, and a total of 200 bands are included to ensure convergence of all computed quantities. After convergence, a phenomenological vertical shift of 0.1~\AA~downwards is imposed on the low-plan Se atoms and the full band structure and interbrand optical matrix elements  recalculated. A very dense k-point mesh (15129 grid points) over the reducible hexagonal Brillouin zone is sampled in the calculations of the k- resolved CD. 

\medskip
\textbf{Supporting Information} \par 
Supporting Information is available from the Wiley Online Library or from the author.

\medskip
\textbf{Acknowledgements} \par 
The authors acknowledge funding from European Union’s Horizon 2020 research and innovation program under grant agreement No 964735 (FET-OPEN EXTREME-IR). The research leading to these results has received partial funding from the the European Union ``Horizon 2020'' research and innovation programme under grant agreement No.881603 "Graphene Core 3", the ANR-16-CE24-0023 "TeraMicroCav" and ANR-2018-CE08-018-05 "BIRDS".

\medskip

%
\bibliographystyle{MSP}
\bibliography{References.bib}

\end{document}


\title{Supplementary Material: \\ Atomic Layer-controlled Nonlinear Terahertz Valleytronics in \\Dirac Semi-metal and Semiconductor \PtSe2}
\maketitle

\justifying

\\
\noindent\author{Minoosh Hemmat}
\author{Sabrine Ayari}
\author{Martin Mi\v{c}ica}
\author{Hadrien Vergnet}
\author{Guo Shasha}
\author{Mehdi Arfaoui}
\author{Xuechao Yu}
\author{Daniel Vala}
\author{Adrien Wright}
\author{Kamil Postava}
\author{Juliette Mangeney}
\author{Francesca Carosella}
\author{Sihem Jaziri}
\author{Qi Jie Wang}
\author{Liu Zheng}
\author{Jérôme Tignon}
\author{Robson Ferreira}
\author{Emmanuel Baudin}
\author{Sukhdeep Dhillon}


\dedication{}
\section{Thickness measurement}
\begin{figure}
    \centering
     \includegraphics{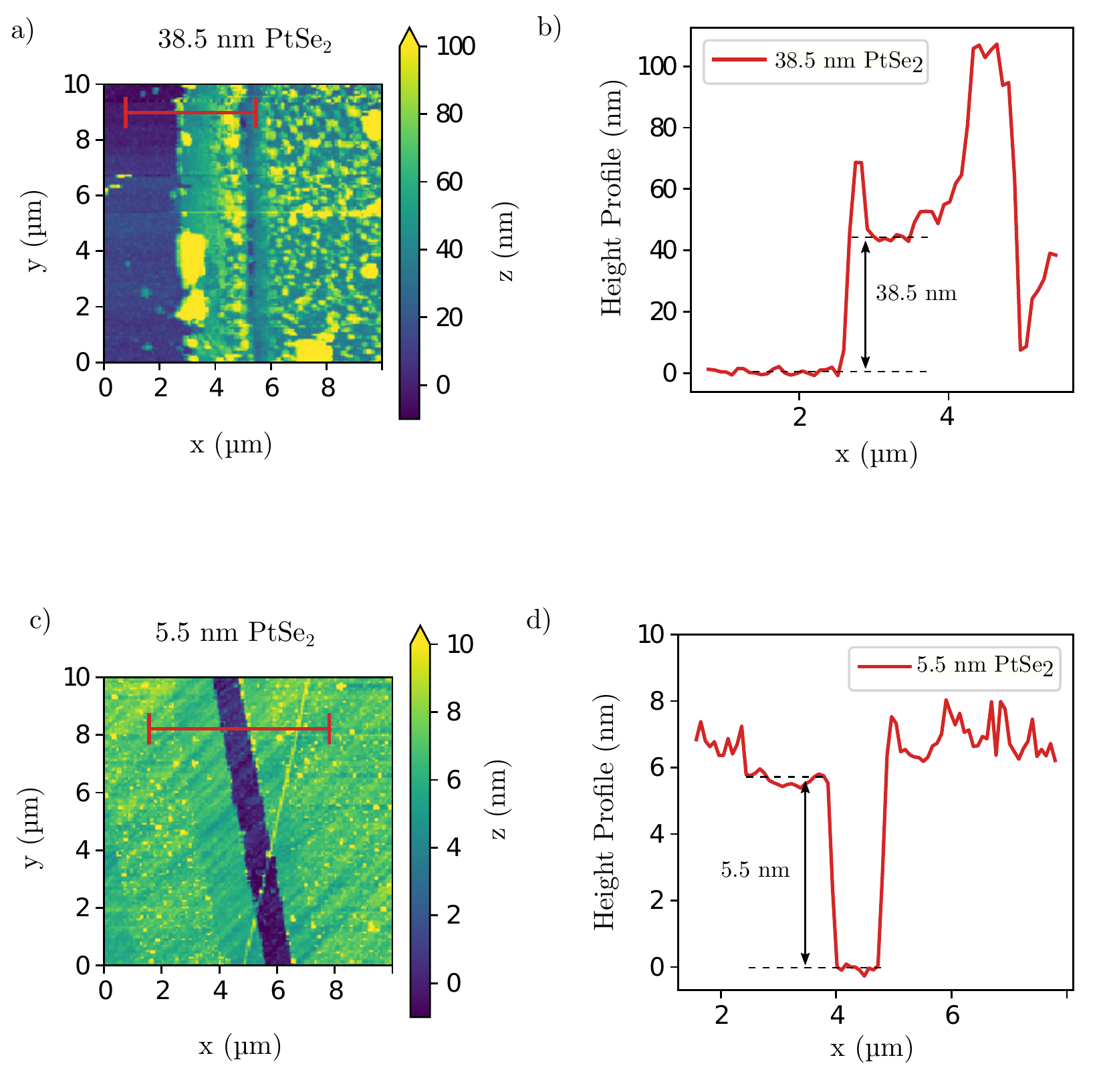}
     \caption{a) and c) AFM images of the 5.5~nm and 38.5~nm \PtSe2 samples. The red lines indicates the extracted sample height profile. b) and d) Height profiles extracted from the respective AFM images.}
     \label{fig:sup_AFM_images}
\end{figure}

The thicknesses of samples were measured using atomic force microscopy (AFM). Figure~\ref{fig:sup_AFM_images} shows AFM images of two \PtSe2 samples and an example of a height profile extracted from each image. 
The reported thickness is obtained by the systematic study of the AFM images. We sample the height difference between the substrate and the \PtSe2 film along the edge of the film. The resulting distribution of height samples is fitted with a Gaussian function to extract the mean thickness value and standard deviation.

\section{Raman measurement}
\begin{figure}
    \centering
     \includegraphics{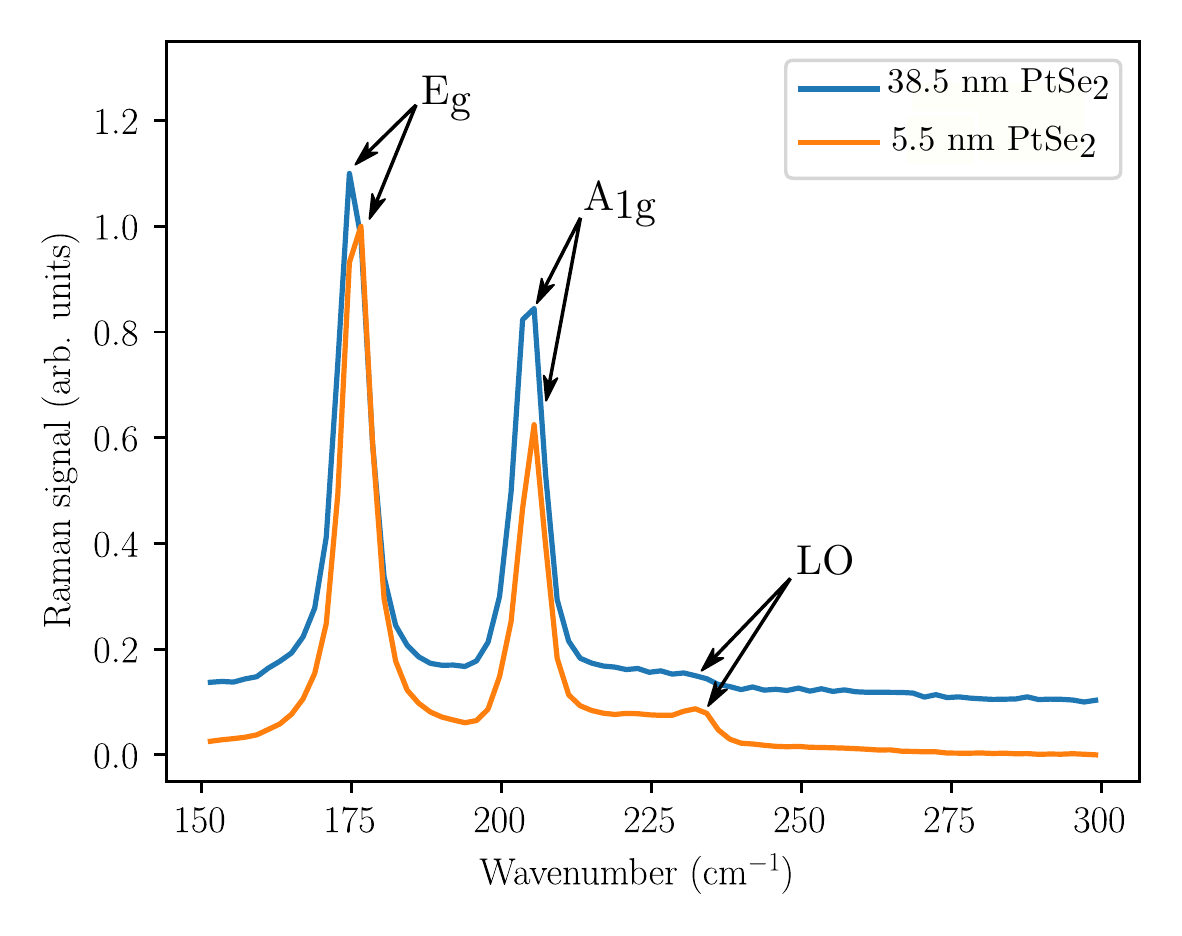}
     \caption{Raman spectra measured on 5.5~nm and 38.5~nm \PtSe2 samples. E$_g$, A$_{1g}$ and LO peaks are identified on each spectrum.}
     \label{fig:sup_Raman_spectra}
\end{figure}

The composition of the samples is checked via Raman spectroscopy using a Renishaw InVia spectrometer with an excitation wavelength at 532~nm and a 100x microscope objective. Figure~\ref{fig:sup_Raman_spectra} shows Raman spectra measured on the two samples that displays the three signature peaks of \PtSe2 : E$_g$ at 177~cm$^{-1}$, A$_{1g}$ at 207~cm$^{-1}$, and LO at 233~cm$^{-1}$ \cite{obrien_raman_2016}.
Peaks observed on the thicker samples are shifted towards lower wavenumbers matching the results present in the literature.

\section{Electronic Properties of Multilayer 1T-PtSe$_2$: A DFT Analysis of Bandstructure, Orbital Projected Partial Density of States and optical matrix elements }

The structural relaxation and electronic properties were performed by Quantum Espresso (QE) code with Density functional theory (DFT) based on Generalized Gradient Approximation (GGA) of the Perdew-Burke-Ernzerhof (PBE) exchange-correlation functional \cite{giannozzi2009quantum,giannozzi2017advanced,perdew1996k}. The norm-conserving pseudopotential was used. The initial lattice constant of 1T-\PtSe2 was taken from the materials project, and was later optimized for the chosen pseudo-potential using the variable cell and quasi-Newton schemes as implemented in the QE package. After convergence tests, the energy cut-off for the plane-wave expansion of the wavefunction is set to 680 eV for all calculations and the Monkhorst-Pack k-point sampling in the Brillouin zone (BZ) is centered with $12 \times 12 \times 12$, $20\times 20\times 20$ , and $32\times 32\times 24$ meshes for geometry optimizations, self-consistent calculation, and projected density of states, respectively. To simulate ML PtSe$_{2}$, we minimize the interaction between the neighboring atomic layers, which are located in subsequent cells. This structure is thus built in such a way that there is a large gap between the neighbors’ atomic layers aligned along the c-axis. In this case, the interaction between the atoms of the successive layers in the c-direction will be eliminated. We performed our calculations by taking the vacuum value equal to 30 Å between two atomic layers along with the c- direction. During optimization process all atoms were free to move in all directions to minimize the internal forces below the threshold $10^{-4}$ Ry/bohr. Bulk \PtSe2, belong to the $D_{3d}^3(P \Bar{3}m1)$ centrosymmetric space group for the global structure and polar point groups $C_{3v}$ and $D_{3d}$ for the Se and Pt sites, respectively \cite{li2016tuning, guo1986electronic,lee1988development}. The optimized lattice parameter is $a=b=$ 3.78~\AA, and $c= $ 5.18~\AA, in agreement with published results \cite{kandemir2018structural,lee1988development,furuseth1965redetermined,villaos2019thickness}. The ML \PtSe2 contains one Pt layer sandwiched between two Se layers, forming a trigonal structure when projected onto the (001) plane. We find that the calculated lattice constant of the pristine ML \PtSe2 in the primitive unit cell is $a=$ 3.72~\AA. which is in good agreement with previous theoretical and experimental results \cite{li2016tuning,wang2015monolayer}. Atomic positions in the ML do not deviate appreciably from the corresponding bulk positions, indicative of strong bonding between the Pt and chalcogen atoms. 

The thickness, defined as vertical distance between uppermost and lowermost Se layers, of 1T-PtSe$_{2}$ ML (2.64~\AA) is also close to the reported value of 2.53~\AA~\cite{wang2015monolayer,kandemir2018structural}. For bandstructure calculation, the bulk lattice constants are used to construct the layered structures starting from monolayer up to 8~ML. Prior to determining the electronic properties, a series of crystal structure relaxations were executed to refine the structural parameters and ensure their convergence. ${P}^{{c} v}(\mathbf{k})=\left\langle\psi_{c \mathbf{k}}|\hat{p}| \psi_{v \mathbf{k}}\right\rangle$ is calculated  from the DFT- Kohn-Sham eigensystem within the non-interaction particules as implemented in  YAMBO code ~\cite{sangalli2019many,marini2009yambo}. In the circular polarization calculations, the whole Brillouin zone was sampled with a very dense k-point grid of 15129 point. 
\begin{figure}
    \centering
     \includegraphics[width=1\textwidth]{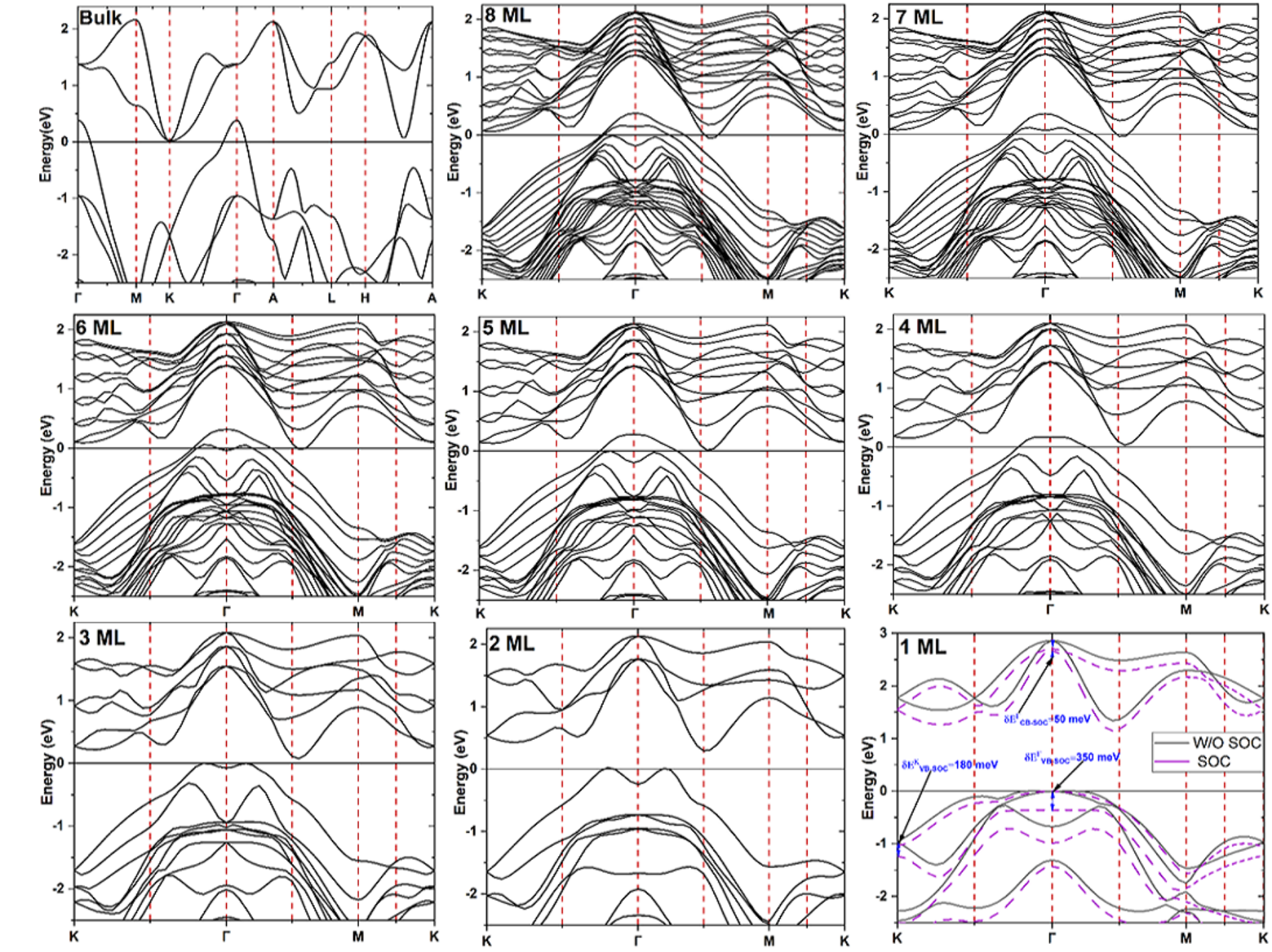}
     \hfil
     \caption{ Electronic band structure of 1T-PtSe$_2$ as a function of thickness was performed, ranging from monolayer to 8-layer, and finally to bulk. The monolayer band structure was presented both with and without the inclusion of spin-orbit coupling (SOC) effects. For clarity, the Fermi level was established as the reference point, indicated by the horizontal solid black line, with a value of zero.}
     \label{fig:band}
\end{figure}

\PtSe2 presents a layer-dependent band structure with dimensional reduction from bulk to monolayer. To understand the effect of the thickness on the electronic structures, we calculate, in Figure~\ref{fig:band}, the band structures of 1~ML to 8~ML  \PtSe2 as well as bulk \PtSe2. The band gap of \PtSe2 shows a sharp decrease as the number of layer increased. This phenomenon is governed by the change of quantum confinement effect and the interlayer interaction through Van der Waals interaction as a function of the number of layer (NL) \cite{kandemir2018structural,ghasemi2020electronic,villaos2019thickness, kuc2011influence,pandey2020layer}. DFT calculations reveal the indirect band nature of 1~ML, 2~ML and 3~ML \PtSe2 with gaps without the spin orbit coupling (SOC) equal to 1.34~eV and 0.3~eV, and 0.08~eV, respectively. From 4~ML, with the increase of NL, the energy level of the valence band maximum (VBM) exceeds that of conduction band minimum (CBM) between $\Gamma$ and M because of the increase in interlayer electronic hybridization, leading to a semiconductor-to-semimetal evolution. In fact, from 8~ML to 4~ML the VBM is located at $\Gamma$ point while the CBM is located in-between the M and $\Gamma$ points. In 2 and 3~ML, both the VBM and CBM, are located in-between M and $\Gamma$ points. Then in 1~ML, the VBM of \PtSe2 goes back again to $\Gamma$ point, which mean that there is minimal to no shift of indirect-to-direct band gap as the thickness decreases for the \PtSe2 structure. Notably in the bulk, the CBM moves from a point between the $\Gamma$ and M to the K point due to the strong interlayer interaction of \PtSe2.

The SOC was included in the DFT calculations by using a full relativistic pseudopotential for ML \PtSe2. We found that relativistic effects in \PtSe2 are significant and lead to strong intrinsic SOC. In fact, with the effect of SOC, double degenerate VBM (fourfold with spin) is separated at $\Gamma$ point by about $\delta E_{VB-SO}^{\Gamma} =$~350~meV. As a result the VBM moves to the brillouin zone (BZ) center and the indirect band gap reduces to 1.15~eV in agreement with published calculations \cite{absor2020spin,kandemir2018structural,kurpas2021intrinsic,gong2020two}. The corresponding spin-orbital spliting of the conduction band at $\Gamma$ point are much smaller, about $\delta E_{CB-SO}^{\Gamma} =$ 50~meV. The orbital degeneracy is also removed at the $K$ point. The energy splitting of the two highest valence bands is $\delta E_{VB-SO}^{K}=$ 180~meV. For materials  without inversion symmetry, the  SOC  result in the lifting of spin degeneracy in the electronic band structure. However, in the case of monolayer PtSe$_2$, our observations reveal that all the bands exhibit a double degeneracy, which is a consequence of the crystal's centrosymmetric property.

\begin{figure}[ht]
    \centering
     \includegraphics[width=1\textwidth]{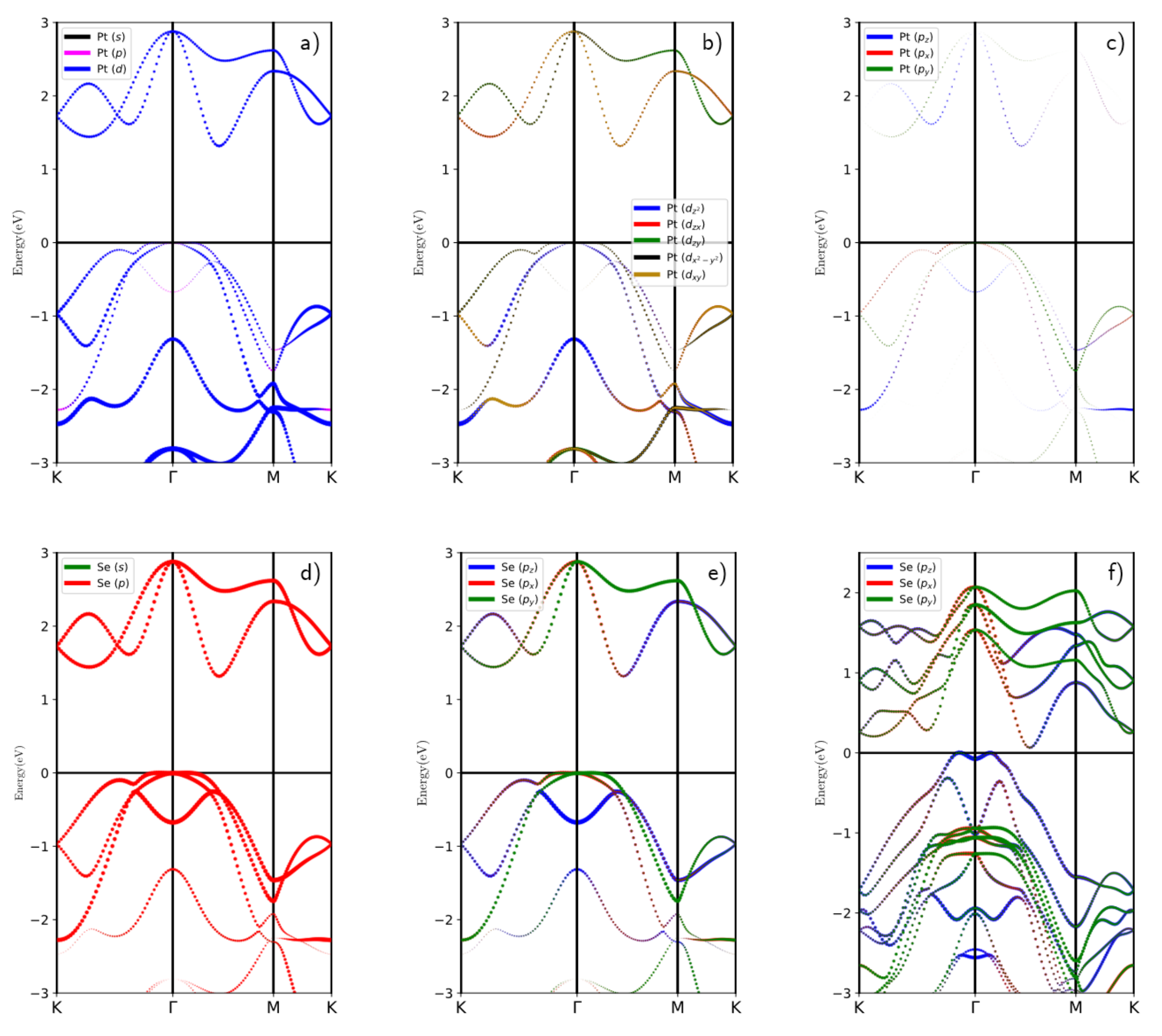}
     \hfil
     \caption{Electronic band structures with orbital projections for ML (a-e) and 3ML \PtSe2 (f)  orbital projections are depicted with varying radii, with the size of the colored circles representing the relative weight of the orbital contributions to the electronic states.}
     \label{fig:dos}
\end{figure}

\noindent In Figure~\ref{fig:dos}, we calculate the orbital projected partial density of states (PDOS) for ML. The calculated PDOS reveals that the valence and conduction bands are mainly composed by d orbitals of Pt and p orbitals of Se atoms, the contribution from s-orbital are absent. In the valence band close to the band gap, the dominant contribution comes from Se p orbitals, with significant admixture of d orbitals from Pt. In the conduction band, the contributions from Pt and Se atoms are almost equal. The VBM, comprised of the $p_{x}$ and $p_{y}$ orbitals of Se atoms. The third valence band is also contributed by the p orbitals of Pt and $p_{z}$ orbital of Se atom. The decrease of the band gap with the number of layer  is a direct result of the exceptional interlayer electronic hybridization of the $p_{z}$ orbital of the Se atom. This phenomenon is demonstrated through the analysis of both ML and multilayer samples, as depicted in Figure~\ref{fig:dos} (e) and (f), respectively.
\begin{figure}
    \centering
     \includegraphics[width=1\textwidth]{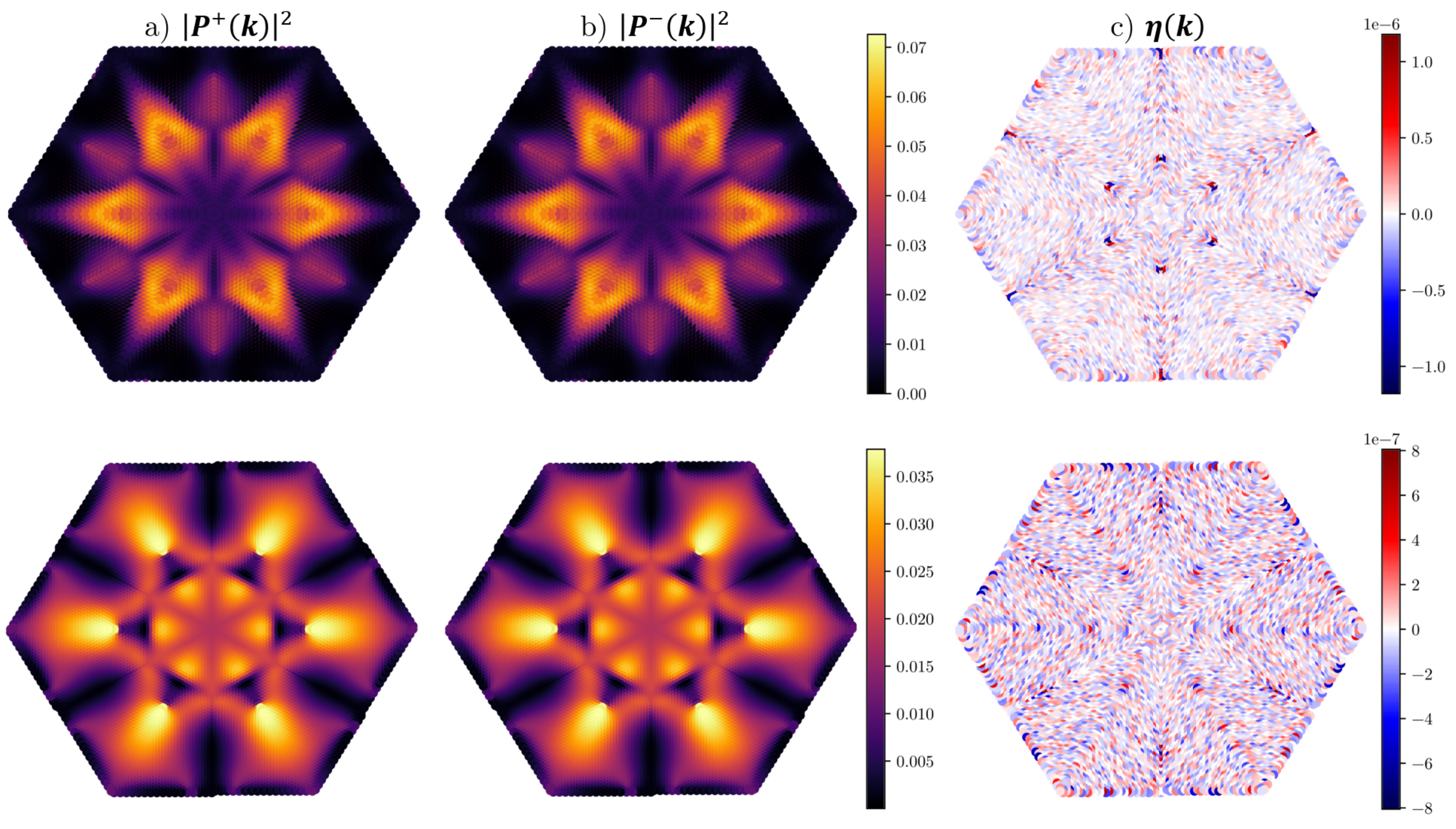}
     \hfil
     \caption{(a,b) the k-resolved direct interband optical matrix elements $\mathcal{P}_{c v}^{\pm}(\mathbf{k})$, and (c) the degree of optical polarization over all energies $\eta(K,\hbar \omega_{c,v} )$ of bulk (upper panel) and ML (lower panel)  of free standing \PtSe2}
     \label{fig:dip}
\end{figure}

Figure~\ref{fig:dip}, show the calculated k-resolved optical matrix element $\left|\mathcal{P}_{c v}^{\pm}(\mathbf{k})\right|^{2}$ and the degree of optical polarization $\eta$ due to the direct interband transition between the top of the valence band (VB) to the bottom of the conduction band (CB) for free standing ML and Bulk \PtSe2. As ML and bulk \PtSe2 has both inversion and time reversal symmetries, all the bands are expected to be doubly degenerate without any net spin polarization. This lead to an equivalent optical transition matrix element $\left|\mathcal{P}_{c v}^{+}(\mathbf{k})\right|^{2} = \left|\mathcal{P}_{c v}^{-}(\mathbf{k})\right|^{2}$.

\section{Formalism for second order emission processes}
In this work, we adapted a formalism to describe the THz emission under pulsed excitation for the general emission and transmission geometry. We can write the nonlinear generation of THz in terms of induced photocurrent and nonlinear conductivity:
\begin{equation}
j_{\lambda}^{(2)}(0,0)=\sigma_{\lambda\nu\eta}^{(2)}(\omega,q)E_{\nu}(q,w)E_{\eta}^*(q,w)
\end{equation}

Where we consider only one incident wave E with frequency w, wavevector q and $\sigma$ is the nonlinear conductivity tensor that can be decomposed to its first order of q:
\begin{equation}
\sigma_{\lambda\nu\eta}^{(2)}(\omega,\boldsymbol{q})=\sigma_{\lambda\nu\eta}^{(2)}(\omega,0)+\Phi_{\lambda\mu\nu\eta}(\omega)q_{\mu}
\end{equation}

The two terms $\sigma$ and $\Phi$ describe photogalvanic and photon drag effects, respectively. Thus we can separate the induced currents by considering these two effects:
\begin{equation}
j_{\lambda}^{(PG)}(0)=\sigma_{\lambda\nu\eta}^{(2)}(\omega,0)E_{\nu}(q,w)E_{\eta}^*(q,w)
\end{equation}
\begin{equation}
j_{\lambda}^{(PD)}(q)=\Phi_{\lambda\nu\eta}^{(2)}(\omega)E_{\nu}(q,w)E_{\eta}^*(q,w)
\end{equation}

These expressions can be further decomposed considering the contributions due to the linear polarized excitations, that correspond to the symmetric terms regarding the exchange of the indexes v and n, and the circular polarized excitations, that correspond to the asymmetric terms for the same indexes. As a result, we obtain:
\begin{equation}
j_\lambda^{(P G)}=\chi_{\lambda \nu \eta}\left\{E_\nu, E_\eta^*\right\}+i \gamma_{\lambda \mu}\left(\vec{E} \wedge \vec{E}^*\right)_\mu \\
\end{equation}
\begin{equation}
j_\lambda^{(\mathrm{PD})}=T_{\lambda \mu \nu \eta} q_\mu\left\{E_\nu, E_\eta^*\right\}+i R_{\lambda \mu \nu} q_\mu\left(\vec{E} \wedge \vec{E}^*\right)_\nu    
\end{equation}

\begin{figure}
    \centering
     {\includegraphics[width=1\textwidth]{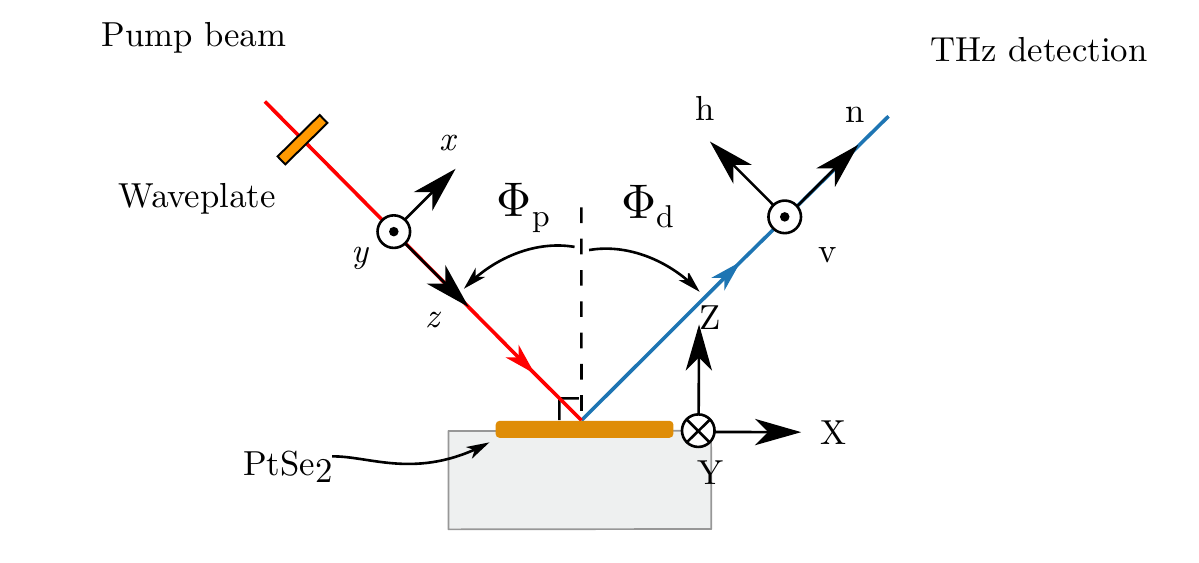}\label{fig:a}}
     \hfil
     \caption{Scheme of the TDS experiment
     in reflection geometry.}
     
     \label{fig:scheme-ref}
\end{figure}

The four terms $\chi_{\lambda \nu \eta}\ $, $\gamma_{\lambda \mu}\ $, $T_{\lambda \mu \nu \eta}$, and $R_{\lambda \mu \nu} $ represent respectively the linear and circular terms of photogalvanic and the linear and circular terms of photon drag effect. They represent phenomenologically the second order nonlinear emissions that we observed from our samples. In order to calculate the THz emission as a function of these four terms, we adapted our measurements by changing the pump polarization and detected THz signal. Figure~\ref{fig:scheme-ref} shows the scheme of the experiment in reflection geometry.

The tensor components are restricted by the symmetry of the crystal structure of \PtSe2 and its environment. This material belongs to the $D_{3d}$ symmetry group. Thus, the non-zero and independent components are:
\begin{equation}
\begin{aligned}
& T_{1} \equiv T_{xxxx}+T_{xxyy}\\
& T_{2} \equiv T_{xxxx}-T_{xxyy}\\
& T_{3} \equiv 2T_{xzxz}\\
& T_{4} \equiv T_{xxzz}
\end{aligned}
\end{equation}
\begin{equation}
\begin{aligned}
& R_{1}\equiv R_{yzx}=-R_{xzy}\\
& R_{2}\equiv R_{xyz}=-R_{yxz} 
\end{aligned}
\end{equation}

The first coefficients are determined exclusively by the symmetry of \PtSe2. The asymmetry that is introduced by the sample environment (The presence of quartz on one side and air the other side) and the geometry of the experiment (the sample is excited from one side) are responsible for the supplementary terms:
\begin{equation}
\chi\equiv \chi_{xxz}= \chi_{yyz}= \chi_{xxx}= \chi_{yzy}
\end{equation}
\begin{equation}
\gamma\equiv \gamma_{xy}= -\gamma_{yx}    
\end{equation}

As depicted in Figure~\ref{fig:scheme-ref}, we transform the xyz basis for the waveplate to XYZ basis of the sample. The pump beam is initially linearly polarized in X axis that corresponds to TM polarization seen by the sample:
\begin{equation}
\vec{E}_{\text {laser }}=\left(\begin{array}{c}
E_x \\
0 \\
0
\end{array}\right)
\end{equation}

The pump beam passing through the waveplate whose fast axis has an angle ($\theta$) with x axis can be described by the following matrix for the HWP and QWP:
\begin{equation}
M_{\lambda / 2}=\left(\begin{array}{ccc}
\cos \left(2 \theta_{\lambda / 2}\right) & * & * \\
\sin \left(2 \theta_{\lambda / 2}\right) & * & * \\
0 & * & *
\end{array}\right) \\
\end{equation}
\begin{equation}
M_{\lambda / 4}=\frac{1}{\sqrt{2}}\left(\begin{array}{ccc}
1-i \cos \left(2 \theta_{\lambda / 4}\right) & * & * \\
-i \sin \left(2 \theta_{\lambda / 4}\right) & * & * \\
0 & * & *
\end{array}\right)
\end{equation}

In the measurements where the polarization of the laser is fixed compared to x axis, we can ignore the coefficients shown by * as that they don’t enter in our experiments.
Afterwards, to know the beam polarization seen by the sample, we should do a rotation on the obtained polarization:
\begin{equation}
M_{x y z \rightarrow X Y Z}=\left(\begin{array}{ccc}
-\cos \left(\phi_p\right) & 0 & \sin \left(\phi_p\right) \\
0 & 1 & 0 \\
-\sin \left(\phi_p\right) & 0 & -\cos \left(\phi_p\right)
\end{array}\right)
\end{equation}

As a consequence, the pump beam with an arbitrary polarization in the basis of the laser is:
\begin{equation}
\vec{E}_{\text {pump}}=M_{x y z \rightarrow X Y Z} M_{W P} \vec{E}_{\text {laser }}
\end{equation}

From Maxwell equations, The generated THz electric field from surface S at distance r by the current density j is in the form:
\begin{equation}
\vec{E}_{\text {emission }}(r)=\frac{i k_{\mathrm{THz}} S}{c}(\vec{n} \wedge \vec{j}) \wedge \vec{n} \frac{e^{i k_{\mathrm{THz}} r}}{r}
\end{equation}

Finally, to calculate this electric field, we need to change the coordinates basis:
\begin{equation}
M_{X Y Z \rightarrow h v n}=\left(\begin{array}{ccc}
-\cos \left(\phi_d\right) & 0 & \sin \left(\phi_d\right) \\
0 & 1 & 0 \\
\sin \left(\phi_d\right) & 0 & \cos \left(\phi_d\right)
\end{array}\right) \\
\end{equation}
\begin{equation}
\vec{E}_{\mathrm{THz}}=\frac{i \sqrt{\Omega} k_{\mathrm{THz}} S}{c}\left(\vec{n} \wedge\left(M_{X Y Z \rightarrow h v n} \vec{j}\right)\right) \wedge \vec{n}
\end{equation}

The THz electric field measured in our TDS experiments in terms of h and v correspond to:
\begin{equation}
E_{\mathrm{THz}, T M}=\vec{E}_{\mathrm{THz}} \cdot \vec{h}
\end{equation}
\begin{equation}
E_{\mathrm{THz}, T E}=\vec{E}_{\mathrm{THz}} \cdot \vec{v}
\end{equation}

In the first TDS experiment in reflection geometry and in the presence of HWP, the THz signal depends on the HWP rotation $\theta_{\lambda/2}$ according to this formula:
\begin{equation}
E_{\mathrm{TH} z, T M}=A+B * \cos \left(4 \theta_{\lambda / 2}\right)
\end{equation}
\begin{equation}
E_{\mathrm{THz}, T E}=C  \sin \left(4 \theta_{\lambda / 2}\right) \\
\end{equation}
\begin{equation}
A \equiv I_{\exp 1} \cos \left(\phi_d\right) \sin \left(\phi_p\right)
\end{equation}
$$
\times\left[\left(T_{x x x x}-T_{x z x z}\right) \cos ^2\left(\phi_p\right)+T_{x x z z} \sin ^2\left(\phi_p\right)+T_{x x y y}+\left(\chi_{x x z}+\chi_{x z x}\right) \cos \left(\phi_p\right)\right. \\
$$
\begin{equation}
B \equiv I_{\exp 1} \cos \left(\phi_d\right) \sin \left(\phi_p\right) 
\end{equation}
$$
\times\left[\left(T_{x x x x}-T_{x z x z}\right) \cos ^2\left(\phi_p\right)+T_{x x z z} \sin ^2\left(\phi_p\right)-T_{x x y y}+\left(\chi_{x x z}+\chi_{x z x}\right) \cos \left(\phi_p\right)\right.
$$
\begin{equation}
 C \equiv-\frac{\sin \left(\phi_p\right)}{2} I_{\operatorname{expl}}\left[\left(T_{\text {yxyx }}+T_{\text {yxxy }}-T_{y z y z}-T_{y z z y}\right) \cos \left(\phi_p\right)+\chi_{y y z}+\chi_{y z y}\right]
 \end{equation}
 
For the same experiment with QWP, the THz electric field is:
\begin{equation}
E_{\mathrm{THz}, T M}=A+B / 2+B / 2  \cos \left(4 \theta_{\lambda / 4}\right)
\end{equation}
\begin{equation}
E_{\mathrm{THz}, T E}=C / 2  \sin \left(4 \theta_{\lambda / 4}\right)+E * \sin \left(2 \theta_{\lambda / 4}\right) 
\end{equation}
\begin{equation}
E \equiv I_{\exp 1}\left[\left(-R_1+R_2\right) \cos \left(\phi_p\right)-\gamma\right] \sin \left(\phi_p\right)
\end{equation}

where the coefficients A, B and C are the same as the experiment with HWP. $I_{\exp 1}$ regroupe the constants and as a free parameter, it allows us to measure the sensitivity of our experiment when we modify the set up to do another measurement. So with these two measurements, the coefficients A, B, C and E could be determined. However, since these coefficients are a mixture of photogalvanic and photon drag effects, we did measurements in transmission geometry to be able to distinguish these effects.
In this geometry, shown in Figure~3b in the paper, $\beta$ changed while the angle between the incident and detection beam $\alpha$ was zero. The THz emission is calculated to be:
\begin{equation}
E_{\mathrm{THz}, T M}=\cos (\beta) \sin (\beta-\alpha)\left[F+G  \sin ^2(\beta-\alpha)+H  \cos (\beta-\alpha)\right]
\end{equation}
\begin{equation}
E_{\mathrm{THz}, T E}=0 
\end{equation}
\begin{equation}
F \equiv-I_{\exp 2}\left(T_{x x z x}-T_{\text {xxxz }}\right)
\end{equation}
\begin{equation}
G \equiv-I_{\exp 2}\left(T_{x x z z}-T_{x x x z}+T_{x x z z}\right)
\end{equation}
\begin{equation}
H \equiv-I_{\exp 2}\left(\chi_{x x z}+\chi_{x x x}\right)
\end{equation}
\begin{equation}
I_{\text {exp } 2}=I_{\text {exp } 1} 2 \cos \left(\phi_d\right) \sin \left(\phi_p\right)\left[-F-G \sin ^2\left(\phi_p\right)-H \cos \left(\phi_p\right)\right] /(A+B)
\end{equation}

Now owing to the coefficients A, B, C, E, F, G and H that we have determined, we can find the tensor components of the nonlinear conductivity:
\begin{equation}
\left\{\begin{array}{l}
T_1=\frac{-2 F}{I_{\exp 2}}-\frac{4}{I_{\operatorname{exp1}}}\left[\frac{H}{2 \sqrt{2}}-C\right] \\
T_2=\frac{-2 F}{I_{\exp 2}}-\frac{2}{I_{\exp 1}}(A-B)-\frac{4}{I_{\exp 1}}\left[\frac{H}{2 \sqrt{2}}-C\right] \\
T_3=\frac{-2 F}{I_{\exp 2}}-\frac{2}{I_{\exp 1}}(A-B)-\frac{8}{I_{\exp 1}}\left[\frac{H}{2 \sqrt{2}}-C\right] \\
T_4=-(F+G) / I_{\exp 2} \\
\chi_{x x z}+\chi_{x z x}=-H / I_{\exp 2} \\
I_{\exp 2}=I_{\exp 1}[-F-G / 2-H / \sqrt{2}] /(A+B) \\
R_2-R_1-\sqrt{2} \gamma=\frac{2 E}{I_{\exp 1}}
\end{array}\right.
\end{equation}

To determine these values in a certain unit and not an arbitrary unit, we first calculated the sensitivity of our TDS setup using the following formula:
\begin{equation}
\Delta P=P_{\text {probe }} 2 \pi \frac{L}{\lambda} n_0^3 r_{41} E_{T H z}
\end{equation}

Where $P_{\text {probe }}=0.306 \mathrm{~W}$ is the power of the pump laser, $\lambda=800 \mathrm{~nm}$ is the wavelength, $L=1 \mathrm{~mm}$ is the thickness of the electro optic crystal ZnTe, $n_0^3=2.99$ is its optic index, $r_{41}=4.0$~pm/V is its nonlinear coefficient and $E_{\text {THz}}$ is the detected THz electric field amplitude. The optic balance has a gain $G_6=5.82 \pm 18 \times 10^9$~ua/W.
\begin{equation}
C_{\text {ua }}=\frac{c}{\sqrt{\Omega} k_{\mathrm{THz}} P_{\mathrm{pump}} Z_0} \frac{1}{P_{\text {probe }} 2 \pi \frac{L}{\lambda} n_0^3 r_{41}} \frac{1}{G_b}
\end{equation}
\begin{equation}
I_{\text {exp } 1}=\frac{\sqrt{\Omega} k_{\mathrm{THz}} P_{\text {pump }} Z_0}{c} P_{\text {probe }} 2 \pi \frac{L}{\lambda} n_0^3 r_{41} G_b
\end{equation}

Finally, we obtained a relation between the measured signal and the normal unit for second order tensors:
\begin{equation}
\frac{1}{I_{\text {expl } 1}}=0.218 \mathrm{~ua}^{-1} \mathrm{mV}^{-1} \mathrm{~s}^{-1} \text { (with }f_{\mathrm{THz}}=1 \mathrm{THz} )
\end{equation}

\bibliographystyle{MSP}
\bibliography{References.bib}